\documentclass[
aps,
prd,
superscriptaddress,
nofootinbib,
floatfix]
{revtex4}
\usepackage{graphicx,
longtable,
color,
bm}
\begin{document}
\title{Neutrino masses and mixing: Entering the era of subpercent precision}
%
\author{        	Francesco~Capozzi}
\affiliation{   	Dipartimento di Scienze Fisiche e Chimiche, Universit\`a degli Studi dell'Aquila, 67100 L'Aquila, Italy}
\affiliation{		Istituto Nazionale di Fisica Nucleare (INFN), Laboratori Nazionali del Gran Sasso, 67100 Assergi (AQ), Italy}
\author{			William Giar\`e}
\affiliation{		School of Mathematical and Physical Sciences, University of Sheffield, Hounsfield Road, 
					Sheffield S3 7RH, United Kingdom}
\author{        	Eligio~Lisi}
\affiliation{   	Istituto Nazionale di Fisica Nucleare, Sezione di Bari, 
               		Via Orabona 4, 70126 Bari, Italy}
\author{        	Antonio~Marrone}
\affiliation{   	Dipartimento Interateneo di Fisica ``Michelangelo Merlin,'' 
               		Via Amendola 173, 70126 Bari, Italy}%
\affiliation{   	Istituto Nazionale di Fisica Nucleare, Sezione di Bari, 
               		Via Orabona 4, 70126 Bari, Italy}
\author{			Alessandro~Melchiorri}
\affiliation{		Dipartimento di Fisica, Universit{\`a} di Roma ``La Sapienza,'' P.le Aldo Moro 2, 00185 Rome, Italy}
\affiliation{   	Istituto Nazionale di Fisica Nucleare, Sezione di Roma~I, 
               		P.le Aldo Moro 2, 00185 Rome, Italy}
\author{        	Antonio~Palazzo}
\affiliation{   	Dipartimento Interateneo di Fisica ``Michelangelo Merlin,'' 
               		Via Amendola 173, 70126 Bari, Italy}%
\affiliation{   	Istituto Nazionale di Fisica Nucleare, Sezione di Bari, 
               		Via Orabona 4, 70126 Bari, Italy}
\begin{abstract}
\medskip
We perform an updated global analysis of the known and unknown parameters of the standard $3\nu$ framework, using data available at the beginning of 2025. The known oscillation parameters include three mixing angles $(\theta_{12},\,\theta_{23},\,\theta_{13})$ and two squared mass gaps, chosen as 
$\delta m^2=m^2_2-m^2_1>0$ and $\Delta m^2=m^2_3-{\textstyle\frac{1}{2}}(m^2_1+m^2_2)$, where the discrete parameter $\alpha=\mathrm{sign}(\Delta m^2)$ distinguishes normal ordering (NO, $\alpha=+1$) from inverted ordering (IO, $\alpha=-1$). With respect to our previous 2021 update, the combination of accelerator, reactor and atmospheric $\nu$ data lead to appreciably reduced uncertainties for $\theta_{23}$, $\theta_{13}$ and $|\Delta m^2|$. In particular, $|\Delta m^2|$ is the first $3\nu$ parameter to enter the domain of subpercent precision (0.8\% at $1\sigma$). We underline some issues about common systematics in combined fits, that might affect (and possibly weaken) this error estimate. 
Concerning oscillation unknowns, we find a relatively weak preference for NO versus IO (at  $2.2\sigma$), for CP violation versus conservation in NO (1.3$\sigma$) and for the first $\theta_{23}$ octant versus the second in NO ($1.1\sigma$). We  discuss the current status and qualitative prospects of the mass ordering hint in the plane charted by the mass parameters 
$(\delta m^2,\,\Delta m^2_{ee})$, where $\Delta m^2_{ee}=|\Delta m^2|+{\textstyle\frac{1}{2}}\alpha(\cos^2\theta_{12}-\sin^2\theta_{12})\delta m^2$, to be jointly measured by the JUNO experiment with subpercent precision.
 We also discuss upper bounds on nonoscillation observables, including the effective $\nu_e$ mass $m_\beta$ in $\beta$-decay, the effective Majorana mass $m_{\beta\beta}$ in $0\nu\beta\beta$ decay, and the total $\nu$ mass $\Sigma$ in cosmology. We report $m_\beta<0.50$~eV 
$(2\sigma)$ from $^3$H data and $m_{\beta\beta}<0.086$~eV ($2\sigma$) from $^{76}$Ge, $^{130}$Te and $^{136}$Xe data, accounting for parametrized nuclear matrix element covariances.  Concerning $\Sigma$, current results show tensions within the standard $\Lambda$CDM cosmological model, pulling $\Sigma$ towards unphysical values and suggesting possible model extensions. We discuss representative combinations of data, with or without augmenting the $\Lambda$CDM model with extra parameters accounting for possible systematics (lensing anomaly) or new physics (dynamical dark energy). The resulting $2\sigma$  upper limits are roughly spread around the bound $\Sigma < 0.2$~eV within a factor of three (both upwards and downwards), with different implications for NO and IO scenarios.   Bounds from oscillation and nonoscillation data are also discussed in the planes charted by pairs of ($m_\beta,\,m_{\beta\beta},\Sigma$) parameters. 
\end{abstract}
\maketitle

\section{Introduction}
\label{Sec:Intro}

Results from solar, atmospheric, accelerator, and reactor neutrino oscillation experiments have established the standard three-neutrino ($3\nu$) framework, where the neutrino states $(\nu_e,\,\nu_\mu,\,\nu_\tau)$ with definite flavor are mixed with neutrino states $(\nu_1,\,\nu_2,\,\nu_3)$ with definite masses $(m_1,\,m_2,\,m_3)$ via a unitary mixing matrix $U_{\alpha i}$ \cite{ParticleDataGroup:2024cfk,PDG1}. 
The current pillars of the $3\nu$ framework are represented by multiple measurements of five parameters: three mixing angles $(\theta_{12},\,\theta_{13},\,\theta_{23})$ governing oscillation amplitudes, and two independent squared mass differences governing oscillation frequencies, that we choose as $\delta m^2=m^2_2-m^2_1>0$ and $\Delta m^2=m^2_3-\textstyle\frac{1}{2}(m^2_1+m^2_2)$. 
Each of these known parameters has been measured at few percent level
in at least two different kinds of 
oscillation experiments, whose combined data analysis allows to test their
consistency, to improve error estimates, and to constrain subleading effects related to
$3\nu$ unknowns. The latter include the unsolved 
$\theta_{23}$ octant ambiguity, the CP-violating phase $\delta$, and the 
neutrino mass ordering, characterized by the discrete parameter 
$\alpha=\mathrm{sign}(\Delta m^2)$, with $\alpha=+1$ for normal ordering (NO) and 
$\alpha=-1$ for inverted ordering (IO). 
Other unknowns, not affecting oscillations, involve the Dirac or Majorana 
nature of neutrinos probed by neutrinoless double beta decay $(0\nu\beta\beta)$,
and the absolute $\nu$ mass scale. The latter is constrained at sub-eV level by 
laboratory $0\nu\beta\beta$ and $\beta$-decay searches for the effective mass parameters 
$m_{\beta\beta}=|\sum_i U_{ei}^2m_i|$ and $m^2_\beta=\sum_i |U_{ei}|^2 m^2_i$, 
respectively, as well as by cosmological probes of the total $\nu$ mass  $\Sigma=m_1+m_2+m_3$. 
Updated overviews of oscillation and nonoscillation neutrino measurements have been presented at 
{\em Neutrino 2024\/} \cite{Nu2024}.

In general, the synergy among different experiments probing the same (known and unknown)
$3\nu$ parameters would be best exploited by combined fits 
performed by the collaborations themselves, especially when their data share common systematics \cite{McDonald24}.  
Recent works in this direction include the joint fit of Super-Kamiokande (SK) atmospheric data and 
T2K accelerator data \cite{Giganti24}, 
as well as the joint fit of T2K and NOvA accelerator data \cite{Wolcott24}, where the effects of various common uncertainties are considered. 
Despite the interest of multi-experiment analyses \cite{Satellite}, 
progress in this direction is slow: obvious sequels (e.g., joint data
fits by more than two collaborations)  and dedicated working groups (as 
those established in collider physics for electroweak precision tests) are not yet envisaged.

Over three decades \cite{Fogli:1993ck}, global analyses by phenomenology groups (external
to collaborations) have provided alternative and useful ways to compare and combine data from a variety of experiments in terms of known and unknown $3\nu$ parameters \cite{Tortola24,Lisi24,Schwetz24}, 
see \cite{deSalas:2020pgw,Esteban:2020cvm,Capozzi:2021fjo,Esteban:2024eli} 
for recent results. In general, such analyses try to construct and reproduce parametric likelihoods ($\chi^2$ maps) as accurately as possible using available data, error estimates and modelling of neutrino 
production, propagation and detection for each oscillation experiment, and 
to extend the statistical analysis to nonoscillation observables as well. 
In some cases, notably for atmospheric neutrino experiments, this task has become 
prohibitively complex for external users, so that related $\chi^2$ maps can only be 
provided by the experimental collaborations. 
With due consideration for unavoidable simplifications or approximations of an external
approach, independent global $\nu$ data analyses have shown remarkable consistency so far, and 
are expected to yield further useful information in the development of the $3\nu$ framework \cite{Schwetz24}. 

In this context, we present an updated global analysis of oscillation and nonoscillation neutrino data, using recent information 
that became available after our previous work in 2021 \cite{Capozzi:2021fjo}, including results
presented at {\em Neutrino 2024\/} \cite{Nu2024} and  up to the beginning of 2025. 
It is a particularly interesting time for an update,
since the known oscillation parameters are entering the era of subpercent precision, 
currently already reached for $|\Delta m^2|$ at face value. Very soon the JUNO reactor experiment alone \cite{Cao24}
is expected to measure with even greater accuracy 
the related parameter $\Delta m^2_{ee}=|\Delta m^2|+{\textstyle\frac{1}{2}}\alpha(\cos^2\theta_{12}-\sin^2\theta_{12})\delta m^2$ 
\cite{JUNO:2015zny},
as well as $\delta m^2$ and $\theta_{12}$ \cite{JUNO:2022mxj}, and to gain increasing sensitivity to the mass ordering options $\alpha=\pm1$ 
\cite{JUNO:2024jaw}. Combinations of accelerator,
reactor and atmospheric data, that have already been showing hints (or tensions) in 
terms of subleading effects, 
are then expected to become even more synergic (or conflicting) in the subpercent
precision era \cite{Schwetz24}. In the next few years, developments in the mass ordering discrimination will thus be especially exciting, together with their cascade effects on the 
likelihood profile of the CP-violating phase $\delta$, 
of the mixing angle $\theta_{23}$ across the two octants, and of the 
absolute mass observables below their upper limits. Concerning the latter, recent
cosmological data seem to push $\Sigma$ towards exceedingly low 
(or even unphysical) values \cite{Elbers24}, that may be suggestive of either experimental tensions  
or of new physics beyond standard cosmology \cite{Gariazzo24}, with potentially interesting---but currently 
uncertain---effects on present bounds or future claims \cite{Jiang:2024viw}. 
Laboratory searches of neutrino mass signals 
are experiencing a steady progress on both $m_\beta$ and $m_{\beta\beta}$, providing independent and complementary mass bounds in the sub-eV range, 
that remains largely to be explored along the allowed NO and IO branches, and that also call for theoretical
progress on nuclear physics aspects; see the related state-of-the-art presentations in \cite{Nu2024}.
Studying the current interplay among all the above oscillation and nonoscillation 
searches adds further motivations for an updated 
global neutrino data analysis. 

Our work is structured as follows. In Sec.~\ref{Sec:Osc} we discuss the 
updated data inputs and the corresponding constraints on known and unknown oscillation parameters, 
shown in terms of projections of regions allowed at
$N_\sigma=\sqrt{\Delta \chi^2}$ standard deviations.  
With respect to \cite{Capozzi:2021fjo}, we obtain improved 
constraints on $\theta_{23}$, $\theta_{13}$ and $|\Delta m^2|$, the latter
being already determined with subpercent accuracy. We discuss
some issues about common systematics in combined fits, that should be better 
understood in view of an increasingly high precision. 
Concerning oscillation unknowns (the mass ordering, the $\theta_{23}$ octant, and the 
CP phase $\delta$), we find slightly weaker hints with respect to \cite{Capozzi:2021fjo}. In particular,
NO is now preferred over IO by $2.2\sigma$. We present the status of this
hint and discuss its possible evolution, using  
the plane charted by the two mass (or frequency) 
parameters that will soon be measured by JUNO,
namely, $\delta m^2$ and 
$\Delta m^2_{ee}=|\Delta m^2|+{\textstyle\frac{1}{2}}\alpha(\cos^2\theta_{12}-\sin^2\theta_{12})\delta m^2$.  
In Sec.~\ref{Sec:Nonosc} we discuss the updated inputs and the corresponding constraints
on $m_\beta$ from tritium $\beta$ decay, and on $m_{\beta\beta}$ from $^{76}$Ge, $^{130}$Te and $^{136}$Xe $0\nu\beta\beta$ data, adapting our previous representation of nuclear matrix element (NME) covariances \cite{Capozzi:2021fjo} to more recent NME evaluations \cite{Menendez24}. Concerning $\Sigma$, we discuss various combinations
of cosmological data within the standard $\Lambda$CDM  model, 
including recent data that tend to push $\Sigma$ 
towards the unphysical region at face value.
We discuss how such constraints on $\Sigma$ are relaxed by augmenting the $\Lambda$CDM 
by physically motivated extra parameters, e.g., to account for the lensing anomaly or
for possible hints of dynamical dark energy. A synopsis of the sub-eV mass bounds emerging from oscillation and nonoscillation data is discussed in the planes charted by pairs of ($m_\beta,\,m_{\beta\beta},\Sigma$) observables.
In Sec.~\ref{Sec:Nonosc} we summarize our findings and briefly comment on perspectives.

\vspace*{-3mm}
\section{Global $3\nu$ analysis of oscillation data and parameters}
\label{Sec:Osc}
\vspace*{-1mm}

We briefly discuss below the updated oscillation inputs with respect to \cite{Capozzi:2021fjo}, covering solar, accelerator, reactor, and atmospheric datasets (Sec.~\ref{Osc1}). We emphasize once more  
some (often overlooked) systematic issues
that may affect combined analyses of accelerator or atmospheric experiments,
with a possible impact on the overall accuracy, e.g., of the $\Delta m^2$ parameter.  
Concerning the methodology, we follow the same approach of \cite{Capozzi:2021fjo},
as originally motivated in \cite{Fogli:2012ua}.
In Sec.~\ref{Osc2}, constraints are first derived from long-baseline (LBL) accelerator, solar and LBL reactor (KamLAND) neutrino data, that represent the minimal dat set sensitive to all the 
$3\nu$ oscillation 
parameters $(\delta m^2,\,|\Delta m^2|,\,\theta_{ij},\,\delta,\,\alpha=\pm1)$. 
Then one adds short-baseline (SBL) reactor data, directly sensitive only to $(|\Delta m^2|,\,\theta_{13})$, but affecting the other parameters via correlations. Finally, by adding
atmospheric data, one gets a further modulation of the oscillation parameter fit,
especially for $(\delta m^2,\,|\Delta m^2|,\,\theta_{23},\,\delta,\,\alpha=\pm1)$.
Oscillation parameters constraints are reported in terms of $N_\sigma=\sqrt{\Delta\chi^2}$ allowed regions,
first for one-parameter projections and then for two-parameter covariances, highlighting the 
impact of various data in constraining knowns and unknowns of $3\nu$ oscillations.
We pay particular attention to the current determination of $\Delta m^2$ at subpercent level and
to the (partly related) mass ordering hints from different data.
In particular (Sec.~\ref{Osc3}) we use the plane charted
by the two frequency parameters ($\delta m^2,\,\Delta m^2_{ee}$) to summarize the pre-JUNO mass-ordering hint 
and to discuss its hypothetical post-JUNO evolution, in terms of either growing 
consistency or possible new tensions among future experimental results. 

\vspace*{-3mm}
\subsection{Oscillation input update and remarks on systematics}
\label{Osc1}

With respect to the previous analysis in \cite{Capozzi:2021fjo}, the solar neutrino input is updated by including 
the SK-IV day and night energy spectra \cite{Super-Kamiokande:2023jbt}, a state-of-the-art standard solar model
\cite{SSM2023,MB22m}, and a very recent reevaluation of the Ga $\nu$ cross section and of its uncertainties \cite{Haxton:2025hye}.
The LBL accelerator input includes the recent spectra of neutrino and antineutrino 
events measured in the appearance and disappearance oscillation channels by the T2K 
\cite{Giganti24,T2K:2023smv,T2K:2023mcm}
and NOvA \cite{Wolcott24,Jargowsky:2024mwc} experiments.
The SBL reactor neutrino input includes the latest results published by Daya Bay \cite{DayaBay:2022orm}
and the complete results recently presented by RENO \cite{RENO:2024msr}. 
For the above classes of experiments it is still feasible---although increasingly complicated---to 
construct parametric likelihoods ($\chi^2$ maps) from the available information, with sufficient accuracy 
for the purposes of a global analysis. 
As usual, we have verified that our single-experiment constraints on oscillation parameters are 
in good agreement with the official ones, whenever a comparison is possible. For atmospheric neutrinos, currently
involving hundred of bins, dozens of systematic uncertainties, and refined statistical separation of 
event classes by flavor proxies, the construction of $\chi^2$ maps based on public information \cite{Fogli:2003th}
has become eventually unfeasible \cite{Capozzi:2018ubv} outside the experimental collaborations. 
We adopt the $\chi^2$ maps officially released for the latest published
atmospheric $\nu$ results by SK \cite{Super-Kamiokande:2023ahc,SKmap} and by IceCube (IC) DeepCore  
\cite{IceCube:2024xjj,ICmap}. We include the official $\Delta\chi^2$ offset between the SK best-fit point for NO (favored)
versus IO \cite{SKmap} (see also \cite{SKatmThesis}). Concerning IC, we take the NO-IO offset as null,
since it has not been officially disclosed \cite{ICmap} (but see \cite{ICThesis} for a related study).    

Following a previous discussion in \cite{Capozzi:2021fjo} (see Secs.~II~B and II~E therein) we emphasize
a few important remarks about systematic errors shared by different
experiments. For solar neutrinos,  common uncertainties (e.g., stemming from the  
the standard solar model) can be included in global analyses via a detailed pull approach \cite{Fogli:2002pt}. 
However, for other classes of
experiments, there is not yet
enough public information to account for shared systematics. This may be a minor
issue if one experiment leads the accuracy in its class (e.g., Daya Bay among SBL reactor experiments) 
but may be of some relevance when combining experiments with comparable statistical power,
such as the LBL accelerator experiments T2K and NOvA. 
Even if they probe different energy ranges, targets and baselines, 
and perform separate near-far detector comparisons, they are expected to share at least some uncertainties 
of the adopted neutrino interaction model(s). Common errors in the reconstruction of the neutrino energy $E$
would represent, e.g., an irreducible uncertainty on $\Delta m^2$ (via the $\Delta m^2/E$ oscillation phase)
in the T2K+NOvA combination. To our knowledge, the coupled effects 
on the joint T2K+NOvA parameter fits have been tentatively evaluated 
in extreme cases (full correlation, no correlation, full anticorrelation), 
but they have not been precisely quantified in a joint analysis pipeline using a common 
model yet \cite{Wolcott24}. Concerning atmospheric $\nu$, 
SK and IC do share significant normalization uncertainties related not only to the
interaction model (applicable to the same water target for overlapping energies) but also
to neutrino flux production model
(up to corrections for the different geomagnetic latitude and depth).
In particular, note that the overall normalization factors (neutrino flux times cross section) 
seem to be pulled in different ways in separate fits: upwards in SK \cite{SKatmThesis} 
and downwards in IC \cite{IceCube:2024xjj}. 
A joint SK+IC fit, enforcing the common pulls to be consistent, might reveal possible effects
on precise parameter estimates, that are lost a priori by simply adding the SK and IC $\chi^2$ maps; see also the
comments in \cite{LisiNPB24}. A joint analysis of simulated data (converging by construction)
seem to suggest that the effect of correlated uncertainties on $\Delta m^2$ is small \cite{Arguelles:2022hrt}.   
 In this context, 
the joint fit of accelerator and atmospheric T2K+SK data \cite{Giganti24,T2K:2024wfn}
is a relevant test case where common interaction and detector model systematics have been officially implemented on real data.
To our understanding, such common systematics 
induce a T2K+SK error component of $3.6\times 10^{-5}$~eV$^2$ 
on $\Delta m^2$ ($1.5\%$) \cite{T2K:2024wfn}, not much smaller than the $E$-scale reconstruction
uncertainty estimated in SK alone ($\sim 1.8\%$) \cite{Super-Kamiokande:2023ahc}, or
the overall accuracy of $\Delta m^2$ quoted by T2K alone ($\sim 2\%$) \cite{Giganti24}.  
We thus surmise that, even with higher statistics, the T2K+SK combined 
error on $\Delta m^2$ might reach an irreducible, common systematic
``floor'' at the level of $\sim 1.5\%$, in the absence of further improvements.
 See also recent studies of interaction model uncertainties
affecting $\Delta m^2/E$ at percent level in T2K \cite{Abe:2024avs} and NOvA \cite{Coyle:2025xjk} event generators.
Summarizing, in the era of subpercent precision, the evaluation of
(partly) irreducible systematics shared by
different experiments deserves further attention by the collaboration themselves, as they
escape control in global fits by external groups.
With the above cautionary remark, we proceed with the discussion of our $3\nu$ oscillation analysis results.

\subsection{$3\nu$ oscillation parameter constraints}
\label{Osc2}

We discuss the constraints  
on the oscillation parameters $(\delta m^2,\,|\Delta m^2|,\,\sin^2\theta_{ij},\,\delta)$ for increasingly rich data sets, in terms of allowed ranges at $N_\sigma$ standard deviations for both NO and IO. 

\begin{figure}[t!]
\begin{minipage}[c]{0.98\textwidth}
\includegraphics[width=0.88\textwidth]{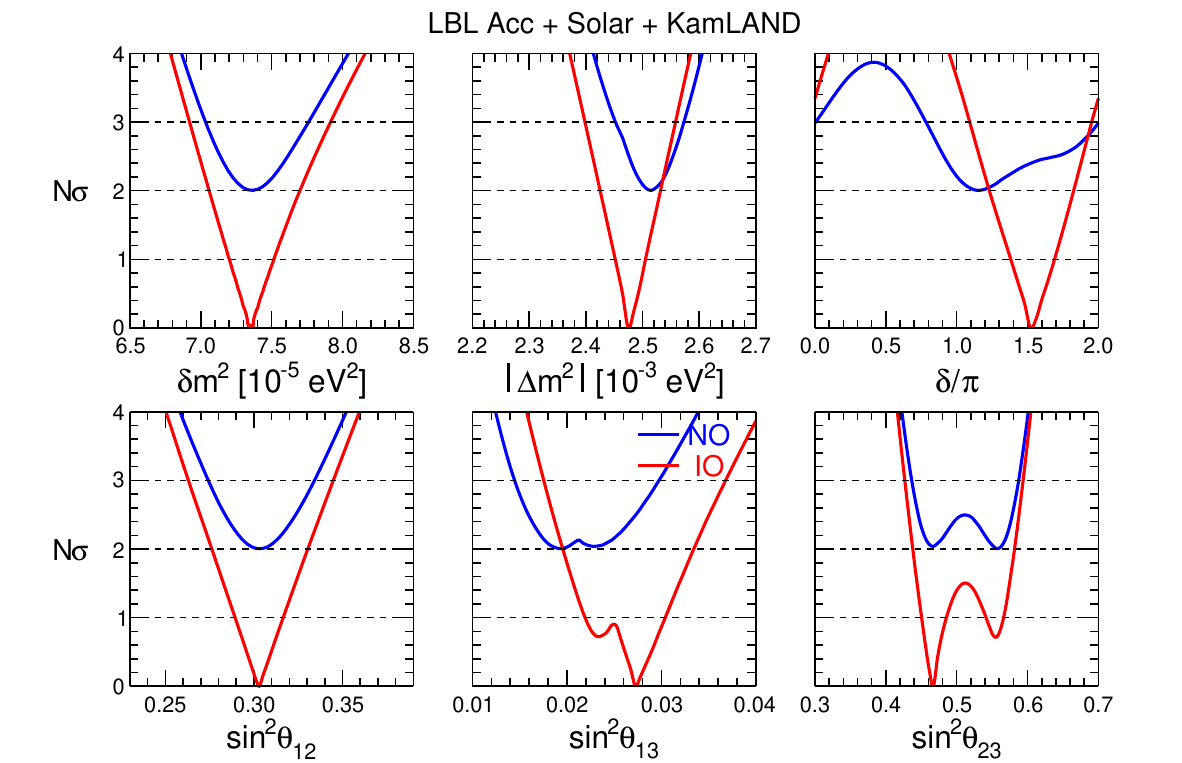}
\caption{\label{Fig_01}
\footnotesize Global $3\nu$ oscillation analysis of long-baseline accelerator, solar and KamLAND $\nu$  data. 
Bounds on the parameters $\delta m^2$, $|\Delta m^2|$, $\sin^2\theta_{ij}$, and $\delta$, for NO (blue) and IO (red).
IO is favored at $2.0\sigma$.   
} \end{minipage}
\end{figure}

Figure~\ref{Fig_01} shows the results for the combination of solar, KamLAND  
and LBL accelerator data for NO (blue) and IO (red), in terms of individual parameters. 
Although the separate T2K and NOvA data provide a slight preference for NO \cite{Giganti24,Wolcott24},
they tend to favor IO in combination \cite{Wolcott24}, at the level of $2\sigma$ in the figure. These findings,
stemming from a mild but persisting tension between T2K and NOvA data, are slightly
more pronounced than in the analysis of previous data \cite{Capozzi:2021fjo}.
The parameters $\delta m^2$ and $\sin^2\theta_{12}$ are rather precisely determined
by solar and KamLAND constraints, with almost linear and symmetrical (i.e., gaussian) 
uncertainties, and no significant best-fit difference between the NO and IO cases.
The parameter $|\Delta m^2|$ is also well costrained,
but with a best-fit value sensitive to the mass ordering. The $\sin^2\theta_{23}$ and 
$\sin^2\theta_{13}$ parameters show two quasi-degenerate minima, due to 
the unsolved $\theta_{23}$ octant ambiguity in  $\nu_\mu$ disappearance, with cascade effects on 
$\theta_{13}$ via $\nu_e$ appearance. As further discussed below, the leading
oscillation amplitude of $\nu_\mu\to\nu_e$ appearance ($\propto \sin^2\theta_{23}\sin^2\theta_{13}$) 
anticorrelates the quasi-degenerate best fits of $\theta_{23}$ and $\theta_{13}$, the latter being 
also noticeably different in NO and IO. Appreciable differences emerge also on the phase $\delta$, that
appears to be largely consistent with CP conservation in NO ($\delta \simeq n \pi$), while being inconsistent with it 
at $>3\sigma$  in IO (with a best fit $\delta \simeq {\textstyle\frac{3}{2}}\pi$, 
see also \cite{Wolcott24}). This rich interplay between known and unknown parameters 
suggests that further direct constraints on the first ones may indirectly affect the second ones, as it indeed occurs.
  
\begin{figure}[b!]
\begin{minipage}[c]{0.98\textwidth}
\includegraphics[width=0.88\textwidth]{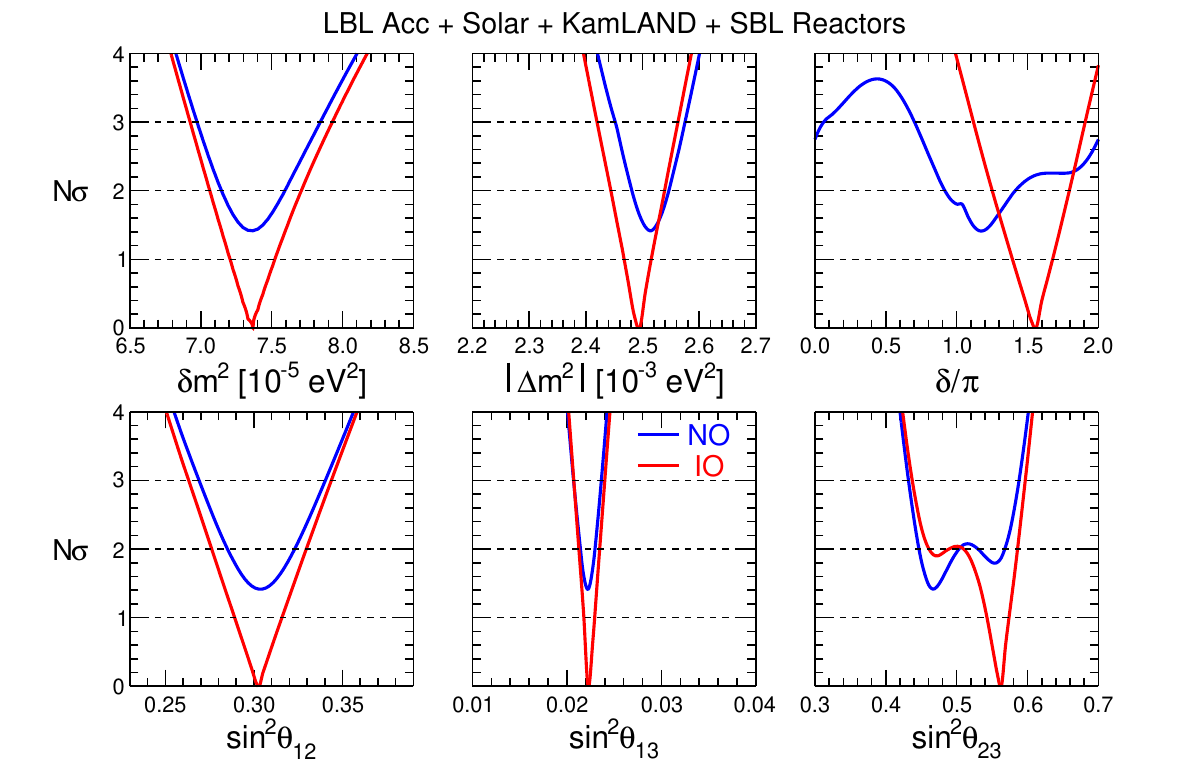}
\caption{\label{Fig_02}
\footnotesize As in Fig.~\protect\ref{Fig_01}, but adding short-baseline reactor $\nu$ data.   
IO is favored at $1.4\sigma$.   
} \end{minipage}
\end{figure}

Figure~\ref{Fig_02} shows the fit including SBL reactor data, which further constrain both
$|\Delta m^2|$ and $\sin^2\theta_{13}$. With respect to the preferred regions in Fig.~\ref{Fig_01}, 
the additional reactor data tend to prefer relatively high values of $|\Delta m^2|$ 
and relatively low values of $\theta_{13}$, which are best reached for NO in the upper $\theta_{23}$ octant. 
As a consequence, the hints on the mass orderings and on $\theta_{23}$ (as well as on $\delta$) also change, despite
the fact that SBL reactor data are not directly sensitive to such unknowns. 
In particular, we find that the combination of solar, KamLAND, LBL accelerator and SBL reactor data 
preference for IO is reduced at $1.4\sigma$. Assuming IO, these data provide a preference
for the second octant of $\theta_{23}$ (at $1.9\sigma$) and a strong indication for CP violation ($\sin\delta<0$ at almost $4\sigma$). Assuming NO, the CP violation trend appears to be much more diluted, and the preferred octant is
even flipped. 

\begin{figure}[t!]
\begin{minipage}[c]{0.98\textwidth}
\includegraphics[width=0.88\textwidth]{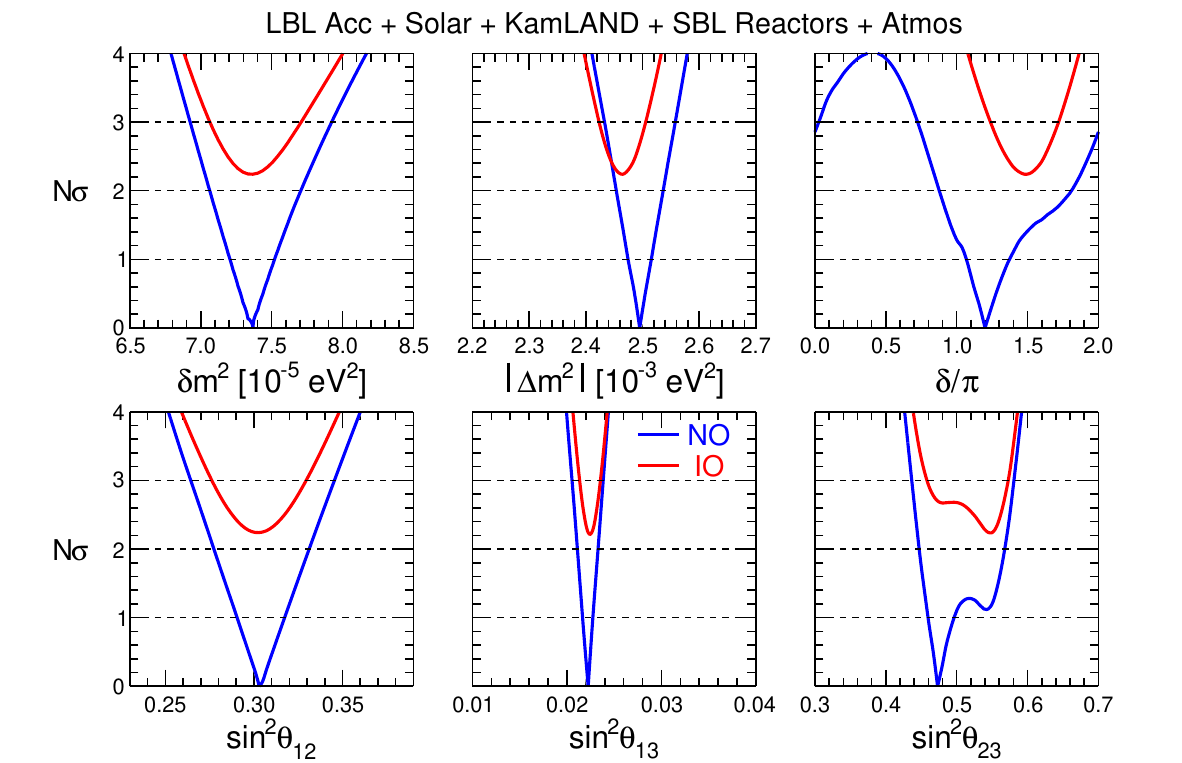}
\caption{\label{Fig_03}
\footnotesize As in Fig.~\protect\ref{Fig_02}, but adding atmospheric $\nu$ data (i.e., with
all oscillation data included). NO is favored at $2.2\sigma$.
} 
\end{minipage}
\end{figure}

Figure~\ref{Fig_03} shows the parameter constraints from the complete oscillation dataset, including atmospheric $\nu$ data from SK and IC. These data add further sensitivity to $|\Delta m^2|$,  $\sin^2\theta_{23}$ and $\delta$. SK data are also particularly sensitive to the mass ordering, preferred to be normal \cite{Super-Kamiokande:2023ahc}. 
This preference wins in the overall combination, with NO being favored at the level of $2.2\sigma$ by all current data 
(while it was $2.5\sigma$ in \cite{Capozzi:2021fjo}).    
The likelihood profiles of $\delta$ and $\theta_{23}$ are also modified by atmospheric data; in particular, for
the NO case, CP conservation is disfavored at $1.2\sigma$ level (while it was $1.6\sigma$ in \cite{Capozzi:2021fjo}), and the first octant is preferred at the
$1.1\sigma$ level ($1.6\sigma$ in \cite{Capozzi:2021fjo}), with respect to the second. 
So the new data, taken altogether, are not reinforcing previous hints about the unknowns, 
partly because of the slight but persisting tension between T2K and NOvA results, as also noted in \cite{Wolcott24,Esteban:2024eli,Chatterjee:2024kbn}. 
The net effect suggests 
caution in the interpretation of current hints on $3\nu$ unknowns.  
This is particularly evident for mass ordering, where the preference for IO by combined LBL accelerator data (Fig.~\ref{Fig_01}) is first reduced by SBL reactor data (Fig.~\ref{Fig_02}) and then flipped to NO by atmospheric data (Fig.~\ref{Fig_03}). 
The current uncertain situation about the $3\nu$ oscillation unknowns
may be contrasted with past converging hints 
about nonzero values for the unknown $\theta_{13}$ 
from the global analysis of solar, KamLAND, atmospheric, and LBL accelerator data
\cite{Fogli:2008jx}, that consistently reached a cumulative statistical above $3\sigma$ \cite{Fogli:2011qn}
before the $\theta_{13}$ discovery and precise measurement by SBL reactor experiments \cite{DayaBay:2012fng,RENO:2012mkc,DoubleChooz:2012gmf}.

Table~\ref{Tab:Synopsis} reports numerically the graphical information of Fig.~\ref{Fig_03}, for 
the separate cases of NO and IO (whose $\chi^2$ difference is given in the last row). 
The known parameters $\delta m^2$ and $\theta_{12}$ are determined at few percent level, with minor changes
as compared with \cite{Capozzi:2021fjo}. Note that we have not included the first published 
results by the SNO+ reactor experiment, as they
constrain $\delta m^2$ with an error larger than in Table~\ref{Tab:Synopsis} 
by a factor of $\sim 6$, and only by assuming a prior on $\theta_{12}$ \cite{SNO:2024wzq}. 
However, SNO+ data expectations on $(\delta m^2,\,\theta_{12})$ are very promising (as indicated by recent preliminary
results \cite{Maneira24})  
and will surpass the current $\delta m^2$ accuracy in a few years, providing   
a relevant input for future global fits. 
With respect to \cite{Capozzi:2021fjo}, from Table~\ref{Tab:Synopsis} we note an appreciable progress in $|\Delta m^2|$, that is 
constrained at $0.8\%$ level at present (it was 1.1\% in \cite{Capozzi:2021fjo}). This is the first oscillation parameter
entering the subpercent precision era, as a result of improved and combined 
results from all SBL reactor, LBL accelerator and atmospheric $\nu$ data. We remark that, as discussed in the previous Sec.~\ref{Osc1}, 
possible effects of common systematics among accelerator or atmospheric experiments  might lead to more conservative fractional uncertainties in a joint fit, that we are unable to evaluate and that should be quantified by experimental collaborations.  
The uncertainty of $\sin^2\theta_{13}$ is  reduced to $2.4\%$ (from $\sim 3\%$ in \cite{Capozzi:2021fjo}). Concerning 
$\sin^2\theta_{23}$, the allowed range is also slightly reduced and, in particular, 
the two quasi-degenerate minima are rather close to each other, differing by only $\sim 15\%$ 
($\sim 25\%$ in \cite{Capozzi:2021fjo}). 
Concerning $\delta$, the constraints are rather similar to those in \cite{Capozzi:2021fjo} within large uncertainties but, as noted,
 with a weaker rejection of CP conservation in NO.
The overall offset between IO and NO is now reduced to $N_\sigma = \sqrt{5.0}= 2.2$ (from $2.5\sigma$ in \cite{Capozzi:2021fjo}).

\newpage

\newpage

Summarizing, in the last few years 
there has been an appreciable progress on three known oscillation parameters ($|\Delta m^2|,\,\theta_{13},\,\theta_{23}$), 
with the first one entering the subpercent precision era. Previous hints about the three oscillation unknowns ($\theta_{23}$ octant, 
CP phase $\delta$, mass ordering) are instead weaker. 
Finally, we note that our global results (Fig.~\ref{Fig_03} and Table~\ref{Tab:Synopsis}) are in good agreement with
ones reported in an independent analysis \cite{Esteban:2024eli}. The agreement would be even better by excluding 
the recent RENO data \cite{RENO:2024msr} appeared after \cite{Esteban:2024eli}; in particular, we would then obtain 
a preference for NO at $2.5\sigma$ as in  \cite{Esteban:2024eli}.

\begin{table}[h!]
\centering
\resizebox{.99\textwidth}{!}{\begin{minipage}{\textwidth}
\caption{\label{Tab:Synopsis} 
Global $3\nu$ oscillation analysis: best-fit values and allowed ranges at $N_\sigma=1$, 2, 3, for  either NO or  IO.  The last column shows the formal  ``$1\sigma$ parameter accuracy,''  defined as 1/6 of the $3\sigma$ range, divided by the best-fit value (in percent). We recall that 
$\Delta m^2=m^2_3-{(m^2_1+m^2_2})/2$ and that $\delta /\pi$ is cyclic (mod 2). Last row:   
$\Delta\chi^2$ offset between IO and NO.
}
\begin{ruledtabular}
\begin{tabular}{lcccccc}
Parameter & Ordering & Best fit & $1\sigma$ range & $2\sigma$ range & $3\sigma$ range & ``$1\sigma$'' (\%) \\
\hline
$\delta m^2/10^{-5}~\mathrm{eV}^2 $ & NO, IO & 7.37 & 7.21 -- 7.52 & 7.06 -- 7.71 & 6.93 -- 7.93 & 2.3 \\
\hline
$\sin^2 \theta_{12}/10^{-1}$ & NO, IO & 3.03 & 2.91 -- 3.17 & 2.77 -- 3.31 & 2.64 -- 3.45 & 4.5 \\
\hline
$|\Delta m^2|/10^{-3}~\mathrm{eV}^2 $ & NO  & 2.495 & 2.475 -- 2.515 & 2.454 -- 2.536 & 2.433 -- 2.558 & 0.8 \\
                                      & IO  & 2.465 & 2.444 -- 2.485 & 2.423 -- 2.506 & 2.403 -- 2.527 & 0.8 \\
\hline
$\sin^2 \theta_{13}/10^{-2}$ & NO & 2.23 & 2.17 -- 2.27 & 2.11 -- 2.33 & 2.06 -- 2.38 & 2.4 \\
                             & IO & 2.23 & 2.19 -- 2.30 & 2.14 -- 2.35 & 2.08 -- 2.41 & 2.4 \\
\hline
$\sin^2 \theta_{23}/10^{-1}$ & NO & 4.73 & 4.60 -- 4.96 & 4.47 -- 5.68 & 4.37 -- 5.81 & 5.1 \\
                             & IO & 5.45 & 5.28 -- 5.60 & 4.58 -- 5.73 & 4.43 -- 5.83 & 4.3 \\
\hline
$\delta/\pi$ & NO & 1.20 & 1.07 -- 1.37 & 0.88 -- 1.81  &   0.73 -- 2.03    & 18 \\
             & IO & 1.48 & 1.36 -- 1.61 & 1.24 -- 1.72  &  1.12 -- 1.83      & 8 \\
\hline
$\Delta \chi^2_{\mathrm{{IO}-{NO}}}$ & IO$-$NO & +5.0   \\ [1pt]
\end{tabular}
\end{ruledtabular}
\end{minipage}}
\end{table}

\newpage 

\begin{figure}[h]
\begin{minipage}[c]{0.85\textwidth}
\includegraphics[width=0.67\textwidth]{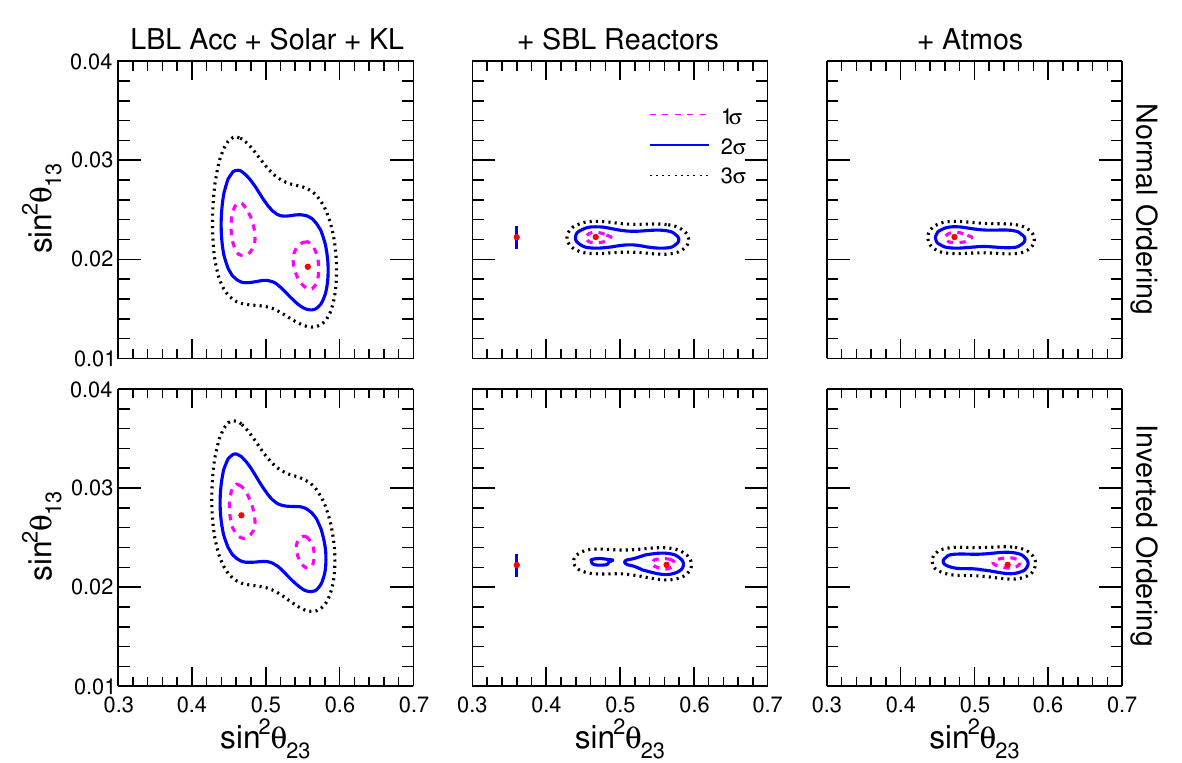}
\vspace*{-2.5mm}
\caption{\label{Fig_04}
\footnotesize Regions allowed in the $(\sin^2\theta_{23},\,\sin^2\theta_{13})$ plane:
Solar + KamLAND + LBL accelerator data (left panels), plus SBL reactor data (middle panels), plus atmospheric data (right panels). Top and bottom panels refer, respectively, to NO and IO as taken separately (i.e., without any relative $\Delta\chi^2$ offset). In the middle panels, the error bars refer to the $\pm2\sigma$  range for $\sin^2\theta_{13}$ arising from SBL reactor data only. 
}
\end{minipage}
\end{figure}

The above single-parameter results and the interplay of different data sets 
can be further understood in terms of selected two-parameter covariances.
Figure~\ref{Fig_04} shows joint constraints on $(\sin^2\theta_{23},\,\sin^2\theta_{13})$
for increasingly rich data sets, in both NO (top) and IO (bottom), in terms of $N_\sigma=\sqrt{\Delta\chi^2}$ isolines, where the 
$\chi^2$ functions are separately minimized for each mass ordering. 
In the left panels, the data are consistent with both octants at $1\sigma$, with a slight preference for the
second octant in NO (first octant in IO). This preference is flipped by adding reactor data (middle panel), that constrain directly
$\theta_{13}$ and indirectly $\theta_{23}$ via correlations. The octant preference is not 
changed by adding atmospheric data (right panel). In any case, the alternative octant is largely allowed within $2\sigma$.

\begin{figure}[b!]
\begin{minipage}[c]{0.85\textwidth}
\includegraphics[width=0.67\textwidth]{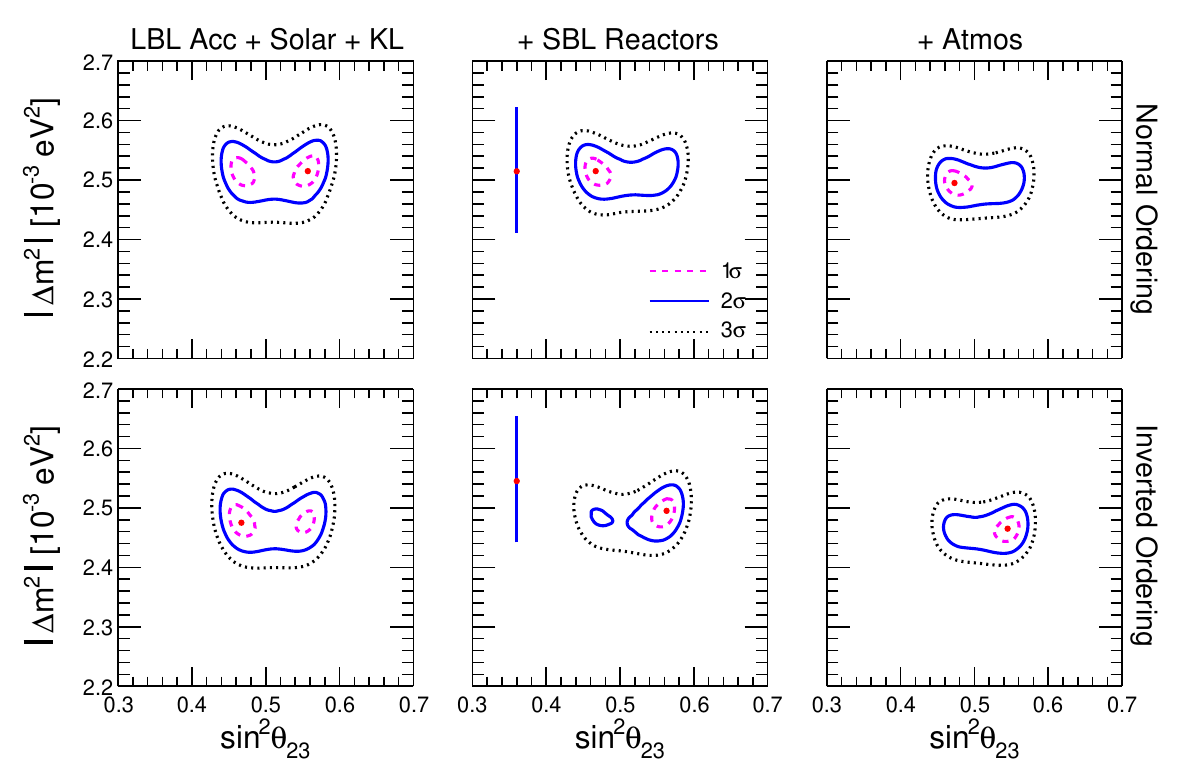}
\vspace*{-2mm}
\caption{\label{Fig_05}
\footnotesize As in Fig.~\ref{Fig_04}, but in the plane 
\textcolor{black}{$(\sin^2\theta_{23},\,|\Delta m^2|)$.}
 The error bars in the middle panels show the $\pm2\sigma$ range for 
 \textcolor{black}{$|\Delta m^2|$} 
 arising from SBL reactor data only. 
}
\end{minipage}
\end{figure}

Figure~\ref{Fig_05} shows the covariance plot for the parameter pair
$(\sin^2\theta_{23},\,|\Delta m^2|)$. 
In this case, notice that the LBL accelerator constraints on $|\Delta m^2|$ (left panels) are in
better agreement with the SBL reactor constraints in NO than in IO (middle panels). This fact 
reduces the joint LBL accelerator preference for IO, as noted in the 
comments to Figs.~\ref{Fig_01} and \ref{Fig_02}. The addition of atmospheric data 
(right panels) reduces the uncertainties on both parameters $(\sin^2\theta_{23},\,|\Delta m^2|)$,
and tends to shift the latter slightly downwards. The slight misalignment of $\Delta m^2$ ranges 
allowed by different data sets is more pronounced in IO than NO, leading to an overall preference 
for the latter in the global fit. The synergy of joint data sets in the search for the true mass ordering \cite{Schwetz24} 
will be further discussed below in Sec.~\ref{Osc3}. 
We emphasize that SBL reactor data, despite having no direct sensitivity to sign($\Delta m^2$) and to $\theta_{23}$,
affect these two variables via covariances in combined fits.

\newpage

\begin{figure}[t!]
\begin{minipage}[c]{0.85\textwidth}
\includegraphics[width=0.65\textwidth]{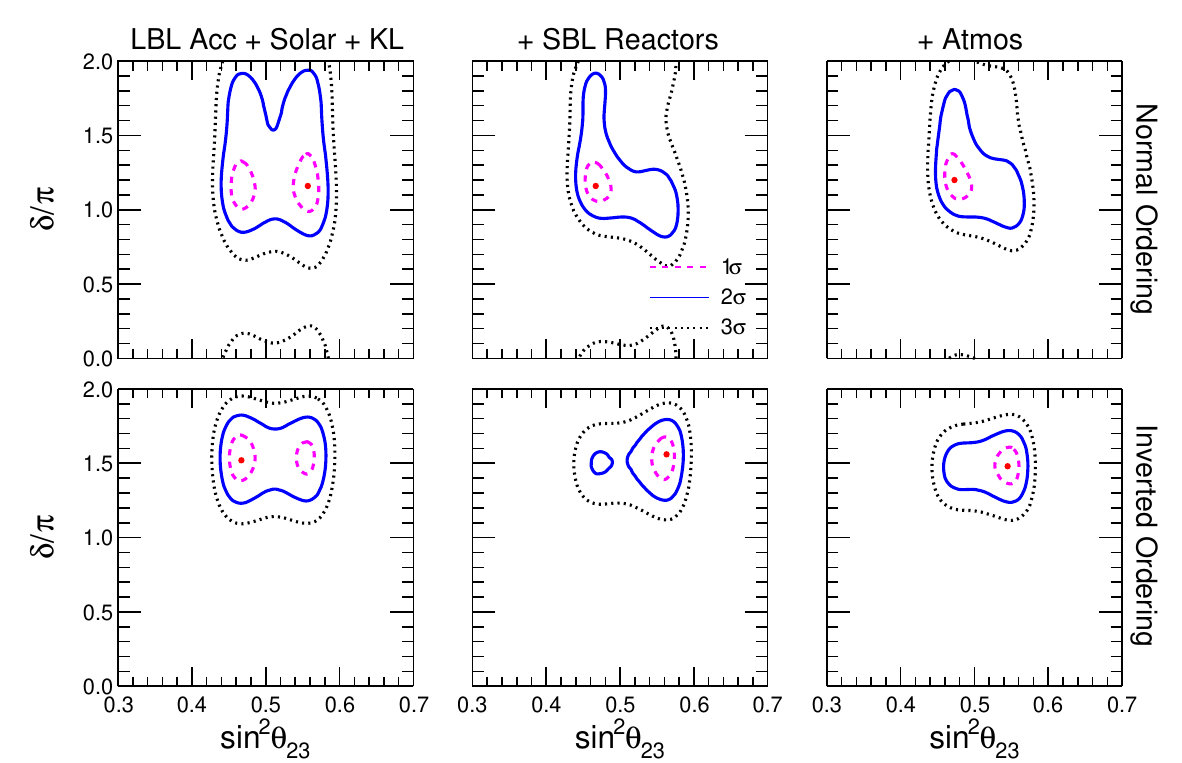}
\vspace*{-2mm}
\caption{\label{Fig_06}
\footnotesize As in Fig.~\ref{Fig_04}, but in the plane $(\sin^2\theta_{23},\,\delta)$.
}
\end{minipage}
\end{figure}

Figure~\ref{Fig_06} shows the covariance of the pair $(\sin^2\theta_{23},\,\delta)$. 
It can be seen that the almost octant-symmetric contours in the left panels 
become rather asymmetric by adding reactor data (middle panels) and then
atmospheric data (right panels). The overall parameter correlation appears to be negative
in NO and negligible in IO, when all data are included; similar results were found previously 
\cite{Capozzi:2021fjo}. These findings imply that, if the octant best fits were hypothetically
flipped, the current slight preference for CP violation 
would be weakened in NO, while it would remain stable in IO. 
This figure illustrates that a weak but interesting interplay already emerges 
among the three oscillation unknowns 
(the CP phase, the $\theta_{23}$ octant and the mass ordering), and that future
data may affect these parameters in subtly correlated ways.

\subsection{Mass ordering hints in JUNO perspective.}
\label{Osc3}

Reactor $\nu$ experiments with a production-detection distance $L$ of tens of kilometers (medium baselines, MBL) can probe
oscillations driven by both $\delta m^2$ and $\Delta m^2$, as well as their interference effects depending on  
the discrete parameter $\alpha=\mathrm{sign}(\Delta m^2)=\pm1$  \cite{Petcov:2001sy}. This accurate oscillometry program 
is going to be carried out by the high-resolution, large-volume JUNO experiment  \cite{Cao24, JUNO:2015zny}, 
whose results are expected to reach subpercent precision
for the ($\delta m^2,\,\Delta m^2$) parameters in less than one year \cite{JUNO:2022mxj}, and $>3\sigma$ sensitivity
to $\alpha=\pm 1$ in less than one decade \cite{JUNO:2024jaw}.

\begin{figure}[t!]
\begin{minipage}[c]{0.87\textwidth}
\includegraphics[width=0.65\textwidth]{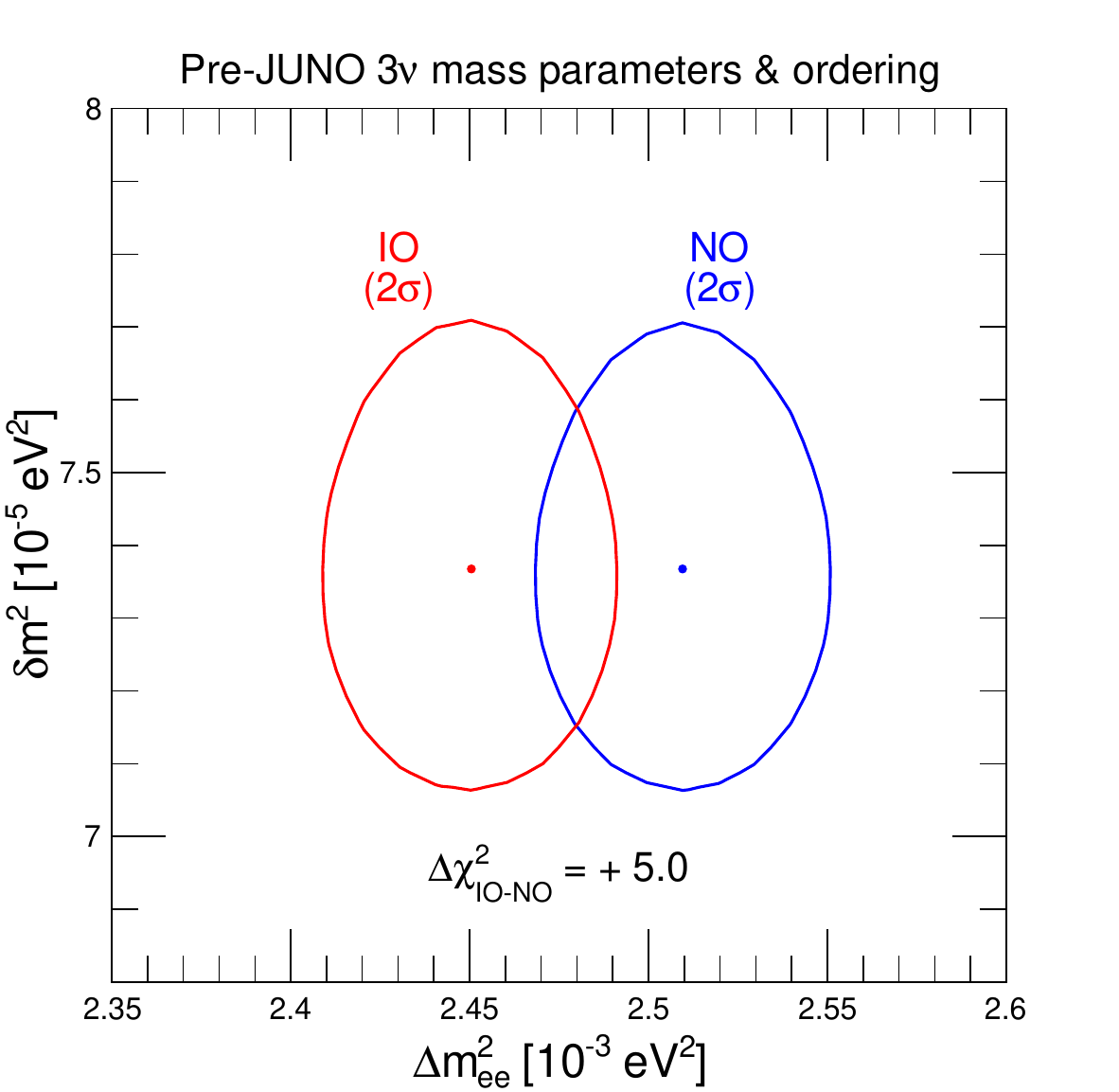}
\vspace*{-2mm}
\caption{\label{Fig_07}
\footnotesize Global $3\nu$ oscillation analysis: Current $2\sigma$ bounds on the mass parameters $(\Delta m^2_{ee},\,\delta m^2)$
that will be directly measured by JUNO, for either NO (blue) or IO (red). The global $\Delta\chi^2$ offset between IO and NO is also reported. The JUNO experiment will provide an independent measurement of both parameters (at subpercent level in less than one year) 
and an independent offset between IO and NO (at $>3\sigma$ in less than one decade). Possible JUNO
outcomes and implications are discussed in the text.  
}
\end{minipage}
\end{figure}

In the context of JUNO (and of reactor $\nu$ experiments in general) 
it is convenient to use, instead of $\Delta m^2$, the frequency parameter
$\Delta m^2_{ee}$ \cite{Nunokawa:2005nx,Minakata:2007tn}. Within JUNO alone, the  sensitivity to $\alpha$
emerges from the relative sign between the leading phase $\Delta m^2_{ee}L/4E$ and a modulation phase 
$\phi$ not scaling as $L/E$
\cite{Minakata:2007tn}. 
Without loss of generality, one can attach the sign $\alpha=\pm 1$ to $\phi$ and take a positive $\Delta m^2_{ee}$,
that in our notation reads (see, e.g., 
\cite{Capozzi:2013psa,Capozzi:2015bpa,Capozzi:2020cxm}):
\begin{equation}
\label{Deltaee}
\Delta m^2_{ee}=|\Delta m^2|+{\textstyle\frac{1}{2}}\alpha(\cos^2\theta_{12}-\sin^2\theta_{12})\delta m^2\ .
\end{equation}
The second term on the r.h.s\ amounts to $\pm 0.015$~eV$^2$, namely, to a $\pm0.6\%$ difference between
$\Delta m^2_{ee}$ and $|\Delta m^2|$. This difference is comparable to both the current $1\sigma$ accuracy of $|\Delta m^2|$
($\pm 0.8\%$, see Table~\ref{Tab:Synopsis}) and to the expected $1\sigma$ accuracy of $\Delta m^2_{ee}$ in JUNO
($\pm 0.8\%$ after just 100 days of operation \cite{JUNO:2022mxj}). Summarizing,
JUNO can contribute to the NO and IO discrimination ($\alpha =\pm 1$) in two ways:
(1) through the relative difference between the phases $\Delta m^2_{ee}L/4E$ and $\phi$ (from JUNO data only);
and (2) through the relative difference between $\Delta m^2_{ee}$ (from JUNO) and $|\Delta m^2|$ or other 
related mass parameters (from non-reactor data).

The second option, namely, the so-called synergy between reactor and non-reactor $\nu$ probes of $\alpha$,
has been explored and illustrated in many ways, using simulated or real data.
An incomplete list includes prospective combinations of 
JUNO-like constraints with LBL accelerator data in disappearance channel 
\cite{deGouvea:2005hk,Minakata:2006gq,Minakata:2007tn,Qian:2012xh,Li:2013zyd,Cabrera:2020ksc,Cao:2020ans,Choubey:2022gzv},
with atmospheric neutrino data in large-volume detectors 
\cite{Blennow:2013vta,IceCube-Gen2:2019fet,KM3NeT:2021rkn} and 
with both accelerator and atmospheric data \cite{Ghosh:2012px,Raikwal:2022nqk}; see also the reviews 
\cite{Qian:2015waa,Qian:2018wid,DeSalas:2018rby,Antonelli:2020uui,CHANDLER:2022gvg}. 
Global analyses of real data have also highlighted the interplay between reactor and non-reactor $\nu$ data
in the emerging hints on mass ordering, see, e.g.,  \cite{Capozzi:2018ubv,Esteban:2016qun} and the above discussion in  Sec.~\ref{Osc2}.
More recently, some studies have combined first hypothetical JUNO outcomes 
with existing (or soon to be expected) rest-of-the-world data
\cite{Cabrera:2020ksc,Forero:2021lax,Parke:2024xre}. The 
mass parameters $(\Delta m^2_{ee},\,\delta m^2)$, directly measurable in JUNO,
appear to be appropriate \cite{Forero:2021lax} for a phenomenological discussion of
near-future perspectives, as discussed below.

In Fig.~\ref{Fig_07} we project the results of our global analysis in the plane charted by the 
$(\Delta m^2_{ee},\,\delta m^2)$ pair, in terms of $2\sigma$ contours ($\Delta\chi^2=4$) for both NO (blue)
and IO (red). The best-fit values of $\Delta m^2_{ee}$ correspond to 
2.450 and $2.510 \times 10^{-3}$~eV$^2$ for IO and NO, respectively. The best-fit separation by $ 0.060\times 10^{-3}$~eV$^2$  
arises from the $|\Delta m^2|$ fit difference ($ 0.030\times 10^{-3}$~eV$^2$) in Table~\ref{Tab:Synopsis}, 
augmented by the intrinsic difference ($\pm 0.015\times 10^{-3}$~eV$^2$)  from Eq.~(\ref{Deltaee}). In Fig.~\ref{Fig_07}
we report, in addition, the  offset $\Delta \chi^2_{\mathrm{{IO}-{NO}}}=5.0$ from
Table~\ref{Tab:Synopsis}. The graphical and numerical results in the figure summarize 
the global information on the neutrino mass parameters, that is
available from ``pre-JUNO'' oscillation experiments.

Let us discuss the implications of possible  JUNO outcomes in this plane.
JUNO is expected to measure the two mass parameters $(\Delta m^2_{ee},\,\delta m^2)$ with 
$2\sigma$ errors smaller than in Fig.~\ref{Fig_07}, in just a few months of operation
\cite{JUNO:2022mxj}. To be precise, the $\Delta m^2_{ee}$ best fit in JUNO will depend also on 
the assumed mass ordering, with a IO value typically higher than the NO value by 
$\sim 0.02\times 10^{-3}\mathrm{\ eV}^2$, depending on fit details \cite{Li:2013zyd,Capozzi:2015bpa,Cabrera:2020ksc,Forero:2021lax}; 
but this difference may well be unresolved in the first JUNO data release, and emerge only with
higher statistics.  
JUNO will also provide its own estimate of
$\Delta \chi^2_{\mathrm{{IO}-{NO}}}$, with a sensitivity that will exceed the value in Fig.~\ref{Fig_07}
in a few years of operation \cite{JUNO:2022mxj}. We make no guess about the exposure 
time and the uncertainties 
associated to the first JUNO data release, and make only qualitative comments 
on the possible central values of the parameters and on the sign of 
$\Delta \chi^2_{\mathrm{{IO}-{NO}}}$ coming from JUNO alone, and in comparison with
pre-JUNO data.  Preliminarily, we note that the currently allowed IO and NO regions in Fig.~\ref{Fig_07} identify three qualitatively different 
$\Delta m^2_{ee}$ ranges at $\sim 2\sigma$ level: an intermediate one where IO and NO results largely overlap ($\Delta m^2_{ee}
\simeq 2.47 - 2.49\times 10^{-3}$~eV$^2$), a left one where IO is preferred over NO (below $\Delta m^2_{ee}
 \sim 2.47 \times 10^{-3}$~eV$^2$) and a right one where NO is preferred over IO (above $\Delta m^2_{ee}
\sim 2.49 \times 10^{-3}$~eV$^2$). When JUNO will present its first results on $\Delta m^2_{ee}$, the position of its
central value in one of these three ranges could  
tell at a glance possible synergic effect with rest-of-the-world constraints in Fig.~\ref{Fig_07}:  A central value located 
in the left (right) range
would imply a marked global preference for IO (NO), while in the intermediate range it would imply a rough 
statistical compatibility with both IO and NO. A possible marked preference for NO would be aligned with the current 
$\Delta \chi^2_{\mathrm{{IO}-{NO}}}=+5.0$  offset in Fig.~\ref{Fig_07}; viceversa, a tension would emerge in IO. 
In addition, JUNO is expected to provide
its own $\Delta \chi^2_{\mathrm{{IO}-{NO}}}$ value, which may or may not have the same sign as the one in Fig.~\ref{Fig_07},
generating another possible source of (mis)alignment among different indications about mass ordering. 
 Global fits will
be useful to gauge the related statistical issues.

We thus draw attention to the different degrees of convergence on mass ordering, that might emerge between pre-JUNO world data 
and first JUNO data. The strongest convergence would be reached for 
$\Delta m^2_{ee} > 2.49\times 10^{-3}$~eV$^2$ and $\Delta \chi^2_{\mathrm{{IO}-{NO}}}>0$ in JUNO,
fully consistent with the pre-JUNO data in favor of NO in Fig.~\ref{Fig_07}.
A puzzling tension would instead emerge for   
$\Delta m^2_{ee} < 2.47\times 10^{-3}$~eV$^2$ and $\Delta \chi^2_{\mathrm{{IO}-{NO}}}>0$ in JUNO,
corresponding to two conflicting indications: one in favor of IO from the
$\Delta m^2_{ee}$ synergy, and another one in favor of NO from the offsets $\Delta \chi^2_{\mathrm{{IO}-{NO}}}>0$. 
Of course, other (mis)alignments may occur, and novel tensions or biases 
may emerge at (sub)percent precision level, providing further motivations  
to revisit the correlated systematic issues mentioned in Sec.~\ref{Osc1}. 
Possible real-data misalignments are usually overlooked in prospective studies, 
where simulated data agree with the ``true'' mass
ordering by construction, and maximally reject the ``wrong'' one. In real cases the synergy may be more nuanced, 
especially when JUNO will start to resolve the mass ordering issue with its own data alone.  

We remark that JUNO's intrinsic sensitivity to mass ordering (i.e., its own $\Delta \chi^2_{\mathrm{{IO}-{NO}}}$) 
depends on the central values of both $\Delta m^2_{ee}$ and $\delta m^2$ 
(and to some extent of $\theta_{12}$), as noted in \cite{Capozzi:2020cxm} and then 
in \cite{Forero:2021lax,JUNO:2024jaw}. In particular, JUNO discriminates better
$\alpha=\pm1$ when the two main oscillation frequencies get closer to each other within current uncertainties, namely, 
when $\delta m^2$ increases or $\Delta m^2_{ee}$ decreases \cite{Capozzi:2020cxm,Forero:2021lax,JUNO:2024jaw}. Therefore we expect that,
for any fixed exposure, JUNO will provide a more
significant offset $\Delta \chi^2_{\mathrm{{IO}-{NO}}}$ for
$(\Delta m^2_{ee},\,\delta m^2)$ values in the upper left part of Fig.~\ref{Fig_07}, and a less significant 
offset in the lower right part. 
Therefore, the exposure needed to solve the mass ordering issue, and the convergence 
among different bits of informations at any given statistical 
significance, appear to depend on the location of the JUNO's best-fit point in the plane $(\Delta m^2_{ee},\,\delta m^2)$,
as well as on the corresponding error bars. Moreover, the (mis)alignment of current bounds on $\delta m^2$ 
with first JUNO results on $\delta m^2$  will provide further relevant information and impact
on other oscillation searches (see, e.g., \cite{Capozzi:2018dat,Denton:2023zwa}).  

In conclusion, Fig.~\ref{Fig_07} summarizes the current information about mass parameters and ordering
in the plane $(\Delta m^2_{ee},\,\delta m^2)$ relevant for JUNO. We have discussed how to 
gauge the qualitative synergy (or misalignment) of this information with first JUNO data in this plane, when
they will be released. More quantitative statements are left to a future global analysis 
including such data, that will be crucial for the development of the $3\nu$ framework.

\section{Global $3\nu$ analysis of absolute mass observables}
\label{Sec:Nonosc}

In this Section we discuss updated bounds on absolute mass observables from nonoscillation data, including:
the effective $\nu_e$ mass $m_\beta$ constrained by $\beta$ decay; the effective Majorana mass $m_{\beta\beta}$ probed
by $0\nu\beta\beta$ decay (if neutrinos are Majorana); and the sum of neutrino masses $\Sigma$ constrained by cosmology.
We use the same notation and conventions of our previous analysis in \cite{Capozzi:2021fjo}, highlighting the main changes in the input data or output results, and the implications for the neutrino mass ordering. A global overview is shown in terms of nonoscillation parameter pairs, following the
graphical format introduced in \cite{Fogli:2004as}.

\subsection{Tritium $\beta$ decay and $m_\beta$ bounds}
\label{Sec:Nonosc1}

The KATRIN experiment has recently released the combined results of the first five 
campaigns of $\beta$-decay spectral measurements with tritium 
\cite{Katrin:2024tvg}. No neutrino mass signal has been found within the uncertainties, 
and an upper limit on the effective mass parameter $m_\beta<0.45$~eV  has been placed at 
90\% C.L.\ (namely, $1.645\sigma$). Upper bounds at different confidence levels, not reported in \cite{Katrin:2024tvg}, 
will appear in supplementary material; relevant ones include $m_\beta<0.35$, 0.50 and 0.61~eV 
at 1, 2 and $3\sigma$ level, respectively, according to preliminary estimates \cite{PrivateKATRIN}.

To a good approximation, the above $n\sigma$ bounds can be reproduced in terms of a simple empirical function $\Delta \chi^2=66.1(m_\beta/\mathrm{eV})^4$
or, equivalently, in terms of an effective $1\sigma$ measurement such as $m^2_\beta = 0\pm 0.123\ \mathrm{eV}^2$.
With respect to \cite{Capozzi:2021fjo} (where KATRIN results were summarized as 
$m^2_\beta = 0.1\pm 0.3\ \mathrm{eV}^2$), the error is appreciably reduced by  a factor $>2$,
and the  $m^2_\beta$ best-fit value is pushed at the border of the physical region. [An allowance for $m^2_\beta<0$ would 
provide a slightly negative best fit \cite{Katrin:2024tvg}.] For graphical purposes,
we shall focus on the corresponding $2\sigma$ upper bound, namely,
\begin{equation}
\label{Boundmb}
m_\beta<0.5\ \mathrm{eV}\ (2\sigma)\ .    
\end{equation}

\subsection{Neutrinoless $\beta\beta$ decay and $m_{\beta\beta}$ bounds}
\label{Sec:Nonosc2}

Searches for $0\nu\beta\beta$ decay signals are of fundamental importance to assess the Dirac or Majorana nature of neutrinos,
see \cite{Gomez-Cadenas:2023vca,Agostini:2022zub} for recent reviews. A decay signal $S_i$ in a candidate isotope (labelled by the
index $i$) would not only prove the Majorana hypothesis but also provide constraints on the effective Majorana neutrino mass $m_{\beta\beta}$. So far, all searched signals are compatible with null values, leading to upper bounds on $S_i$ and $m_{\beta\beta}$. 

In order to quantify such bounds, we follow the methodology of \cite{Capozzi:2021fjo} and define the 
signal strength as 
\begin{equation}
\label{Signal}
S_i=1/T_i=G_i|M_i|^2m^2_{\beta\beta}\ ,
\end{equation}
 where $T_i$ is the decay half-life, $G_i$ is the phase space factor, 
and $M_i$ the nuclear matrix element (NME). 
To a good approximation, 
the experimental likelihood profile of $S_i$ can be parametrized through a quadratic form \cite{Capozzi:2021fjo}
\begin{equation}
\label{DeltaS}
\Delta \chi^2(S_i)=a_i S_i^2+b_iS_i+c_i \ .
\end{equation}

Table~\ref{tab:abc}  reports our numerical evaluation of the coefficients $(a_i,\,b_i,\,c_i)$ 
for both separate and combined bounds on $S_i$, for $i={}^{76}$Ge, $^{130}$Te and $^{136}$Xe,
updating an analogous Table previously reported in \cite{Capozzi:2021fjo}. In particular,
our input data include the final $^{76}$Ge results 
from GERDA \cite{Agostini:2020xta} and MAJORANA \cite{Majorana:2022udl}, the latest $^{130}$Te results from CUORE \cite{CUORE:2024ikf},
and the complete $^{136}$Xe results from KamLAND-Zen \cite{KamLAND-Zen:2024eml} and EXO-200 \cite{Anton:2019wmi}. 

\begin{figure}[b!]
\begin{minipage}[c]{0.9\textwidth}
\includegraphics[width=0.7\textwidth]{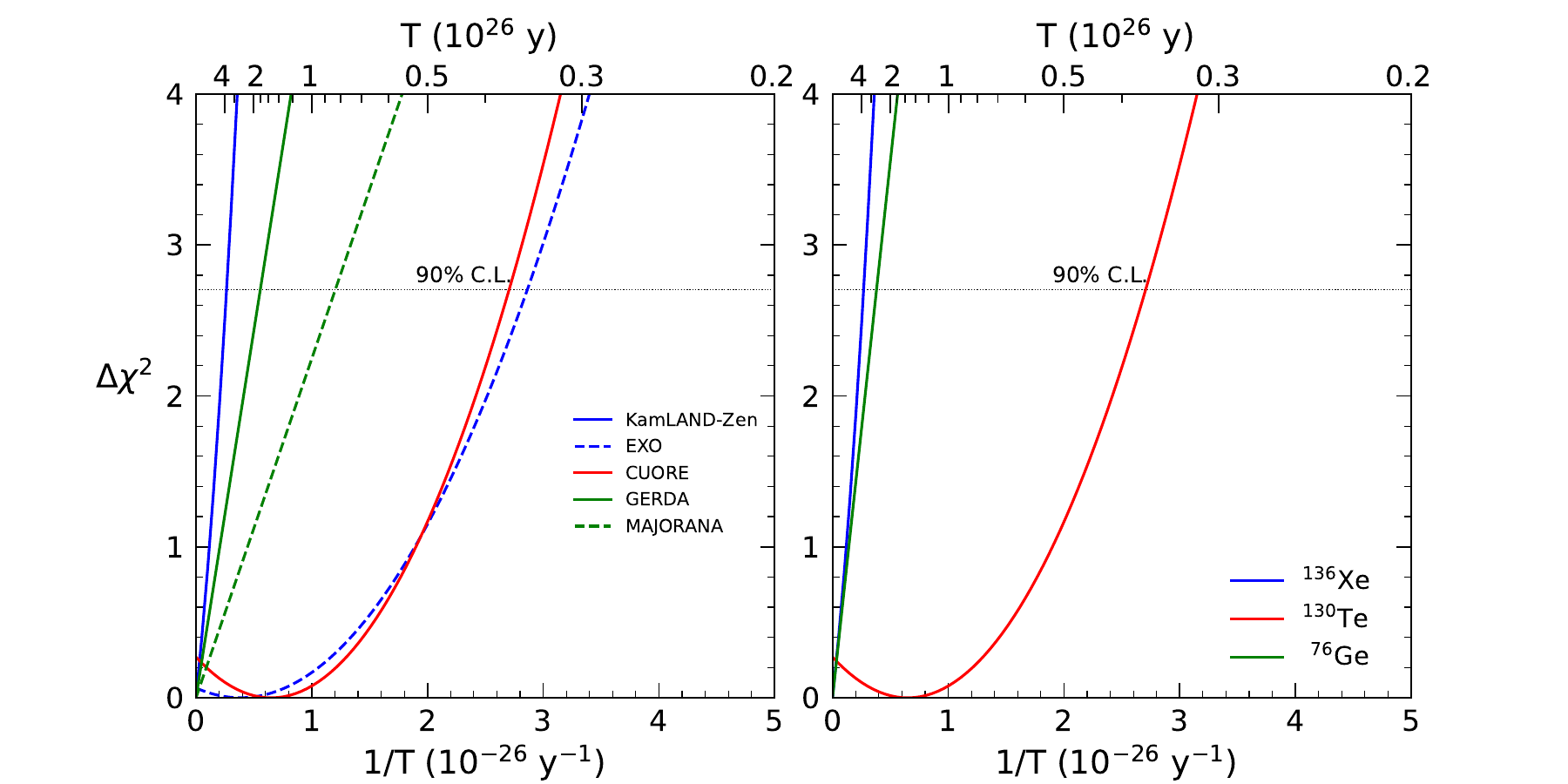}
\caption{\label{Fig_08}
\footnotesize 
$0\nu\beta\beta$ decay: $\Delta\chi^2$ functions
in terms of the half life $T$ (top abscissa) and of the signal strength $S=1/T$ (bottom abscissa). 
Left and right panels: separate experiments and their combinations for the same nuclide, respectively.
Dotted horizontal lines intersect the curves at 90\% C.L.}
\end{minipage}
\end{figure}

Figure~\ref{Fig_08} shows the $\Delta\chi^2$ functions
in terms of $1/T=S$ (bottom abscissa) and of $T$ (top abscissa), for both
separate experiments (left panel) and their combinations for the same nuclide (right panel).
The EXO-200 and CUORE experiments show a small nonzero signal at best fit, that is not statistically significant. 
The dotted horizontal line sets the 90\% C.L.\ upper limits
$T_{90}$ reported in Table~\ref{tab:abc}.

\begin{table}[t!]
\centering
\resizebox{.95\textwidth}{!}{\begin{minipage}{\textwidth}
\caption{\label{tab:abc} 
$0\nu\beta\beta$ decay: Details of the adopted parametrization $\Delta\chi^2(S_i)=a_i\, S_i^2 + b_i\,S_i + c_i  
$ for the signal strength $S_i=1/T_i$, 
in units of 10$^{-26}$~y$^{-1}$. The first two columns report the nuclide and the
name of the experiment(s). The next three columns report our estimated 
coefficients $(a_i,\,b_i,\,c_i)$, taken either separately (upper six rows) or in combination
for the same nuclide (lower three rows).  The fifth column reports our 90\% C.L.\ 
($\Delta\chi^2=2.706$) half-life limits $T_{90}$ in units of $10^{26}$~y, to be compared with the official experimental limits
in the sixth column, as quoted in the references in the last column. 
}
\begin{ruledtabular}
\begin{tabular}{rlrrrccc}
Nuclide 	& Experiment(s) 			& $a_i~~~$ 	& $b_i~~~$ 	& $c_i~~~$ 	& $T_{90}/10^{26}\,\mathrm{y}$ & $T_{90}$ (expt.)
										& References \\ 
\hline
$^{76}$Ge	& GERDA						& 0.000 	& \textcolor{black}{4.867} 	& 0.000 	& 1.800	& 1.8
										& \cite{Agostini:2020xta}\\
$^{76}$Ge	& MAJORANA					& 0.000		& \textcolor{black}{2.246} 	& 0.000		& 0.830 & 0.83 
										& \cite{Majorana:2022udl}\\
$^{130}$Te	& CUORE						& 0.640		& $-0.832$ 	& 0.270		& 0.370 & 0.38 
										& \cite{CUORE:2024ikf}\\
$^{136}$Xe	& KamLAND-Zen 				& 10.664	& 7.412		& 0.000		& 3.781 & 3.8 
										&\cite{KamLAND-Zen:2024eml} \\
$^{136}$Xe	& EXO-200					& 0.440		& $-0.338$ 	& 0.065		& 0.350 & 0.35 
										& \cite{Anton:2019wmi}\\
\hline
$^{76}$Ge	& GERDA + MAJORANA			& 0.000		& 7.074		& 0.000		& 2.070 & \textemdash 
										& This work \\
$^{130}$Te	& CUORE	(same as above)		&  0.640		& $-0.832$ 	& 0.270		& 0.370 & 0.38 
										& \cite{CUORE:2024ikf} \\
$^{136}$Xe	& KamLAND-Zen + EXO-200			
										& 11.104		& 7.074		& 0.000		& 3.718 & \textemdash 
										& This work
\\
\end{tabular}
\end{ruledtabular}
\end{minipage}}
\end{table}

In order to translate the experimental constraints on $S_i$ [Eq.~(\ref{DeltaS})] into constraints on $m_{\beta\beta}$ 
[Eq.~(\ref{Signal}], one must account for the large spread of 
theoretical NME calculations in each isotope \cite{Menendez24}  as well as for their significant correlations between pairs 
of isotopes \cite{Lisi:2022nka}. We follow the methodology of \cite{Capozzi:2021fjo}, based on the only
work where---to our knowledge---a covariance matrix for the relative errors affecting the $|M_i|$ has been estimated 
\cite{Faessler:2008xj}. In particular, the NME central values $|\bar M_i|$ are embedded into the nuclear factors
\begin{equation}
q_i = G_i |\bar M_i|^2
\end{equation}
(in units of $10^{-26}\,\mathrm{y}^{-1\,}\mathrm{eV}^{-2}$)
while their large multiplicative uncertainties (i.e., the errors on $\log |M_i|$) 
are embedded into a dimensionless covariance matrix $\sigma_{ij}$
based on \cite{Lisi:2022nka}. A $\chi^2(m_{\beta\beta})$ function 
is then built to account for both experimental errors
and correlated theoretical uncertainties \cite{Capozzi:2021fjo}. 

\begin{table}[t!]
\centering
\begin{minipage}{.32\textwidth}
\caption{\label{tab:NME} \footnotesize 
$0\nu\beta\beta$ decay: Adopted input for the NME central values and uncertainties,
in terms of coefficients $q_i=G_i|\bar M_i|^2$ and covariance matrix $\sigma_{ij}$; see the text for details.}
\begin{ruledtabular}
\begin{tabular}{cc|ccc|c}
$i$ & Nuclide  &  & $\sigma_{ij}$ & & $q_i$ \\
\hline
1 & $^{76}$Ge  & 0.0790 &        &        &  34.0 \\
2 & $^{130}$Te & 0.0920 & 0.1325 &        & 157.5 \\
3 & $^{136}$Xe & 0.0975 & 0.1437 & 0.1858 &  73.1 \\
\end{tabular}
\end{ruledtabular}
\end{minipage}
\end{table}

\begin{figure}[b!]
\begin{minipage}[c]{0.9\textwidth}
\includegraphics[width=0.7\textwidth]{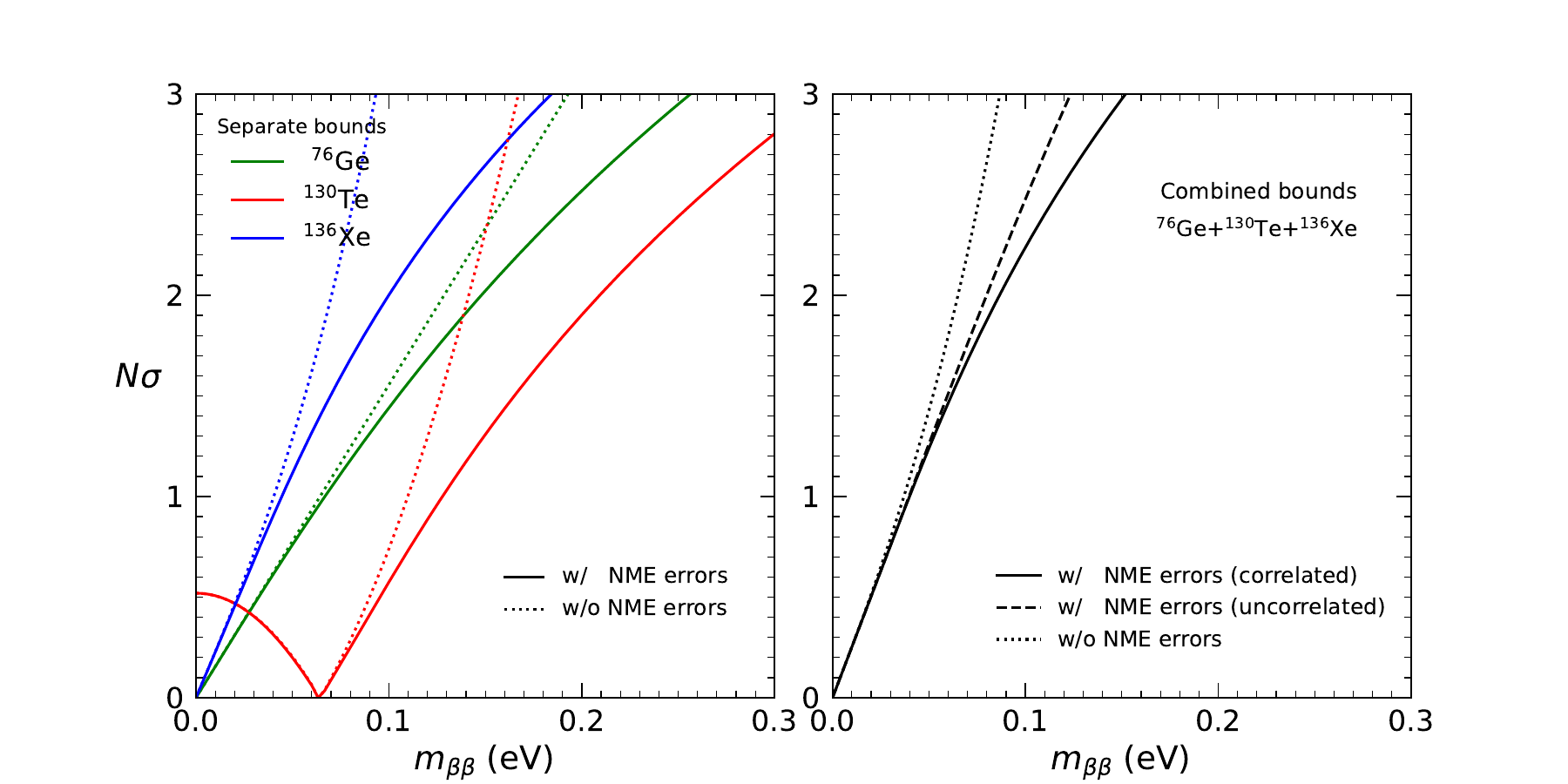}
\caption{\label{Fig_09}
\footnotesize 
$0\nu\beta\beta$ decay: 
Bounds on $m_{\beta\beta}$, expressed in terms of $N_\sigma=\sqrt \Delta\chi^2$. 
The left and right panels refer, respectively, to separate and combined bounds from the three nuclides,  
with (solid) or without (dotted) NME uncertainties. In the right panel, 
the case with uncorrelated uncertainties is also shown (dashed).}
\end{minipage}
\end{figure}

With respect to \cite{Capozzi:2021fjo}, we lower the central values $q_i$ by factors 0.60 and 0.75 
for $^{64}$Ge and $^{130}$Te, respectively, 
to account for the fact that 
recent NME evaluations for these isotopes \cite{Menendez24} 
are generally lower than older estimates considered in \cite{Faessler:2008xj}. 
Since the rescaling has no effect on multiplicative errors, 
we keep the same covariance matrix $\sigma_{ij}$ as in \cite{Capozzi:2021fjo}.
For the sake of completeness, in Table~\ref{tab:NME}
we report the numerical values of $q_i$ and $\sigma_{ij}$ adopted herein.%
\footnote{We also report a misprint in Table~III of \cite{Capozzi:2021fjo}, where the $\sigma_{22}$ entry
was incorrectly typed as 0.0135 instead of 0.1325.}

We have checked 
that, with such inputs, most of the recent NME values compiled 
in the reviews \cite{Menendez24,Gomez-Cadenas:2023vca,Agostini:2022zub} are covered within 
our $\pm2\sigma$ ranges, and a few outliers within $\pm3\sigma$. We thus
surmise that our covariance matrix can conservatively cover 
the large spread of recent NME evaluations. 
Of course, it would be highly desirable to revisit the whole issue of 
NME covariances, in view of recent theoretical developments 
(e.g., {\em ab initio\/} methods) \cite{Menendez24}
that, however, are still focused on single NME uncertainties and not yet on pairwise correlations.
New covariance estimates would lead to an updated Table~\ref{tab:NME} with better motivated entries, without necessarily altering the general
methodology  adopted in \cite{Capozzi:2021fjo} and herein.  

Given the experimental input in Table~\ref{tab:abc} and the adopted theoretical NME input in Table~\ref{tab:NME},
we can work out quantitative bounds at statistical level $N_\sigma = \sqrt {\chi^2(m_{\beta\beta})}$. 
Figure~\ref{Fig_09} shows the resulting bounds on $m_{\beta\beta}$ from 
each separate isotope (left panel) and their combination (right panel). The dominant 
bounds are placed by $^{136}$Xe results, followed by $^{76}$Ge and $^{130}$Te constraints.
The solid lines include experimental and theoretical uncertainties;
the dotted lines only the experimental ones. In addition, the dashed line in the right
panel shows the effect of including the theoretical uncertainties without correlations. 
The most conservative bounds are obtained by including both NME errors and their correlations, leading to the
following $2\sigma$ upper limit:
\begin{equation}
\label{Boundmbb}
m_{\beta\beta}<0.086\ \mathrm{eV}\ (2\sigma)\ .
\end{equation}

The above limit should be compared with the previous analogous estimate $m_{\beta\beta}<0.11$~eV  ($2\sigma$) \cite{Capozzi:2021fjo}, 
and with earlier (less refined) $2\sigma$ estimates, such as $m_{\beta\beta}<0.14$~eV \cite{Capozzi:2020} and $m_{\beta\beta}<0.18$~eV \cite{Capozzi:2017ipn}, reflecting the overall progress on the sensitivity to  $m_{\beta\beta}$ 
(by a factor of $\sim 2$ in the last few years). A reduction of the NME (co)variances would also
be beneficial to improve the bounds, as shown by the results in Fig.~\ref{Fig_09}.


\subsection{Cosmology and $\Sigma$ bounds}
\label{Sec:Nonosc3}

The year 2024 marked a significant turning point for neutrino cosmology. Assuming a minimal $\Lambda$CDM+$\Sigma$ extended cosmological model, pre-2024 cosmological constraints on $\Sigma$ ranged from the conservative 95\% confidence level (CL) limit of $\Sigma \lesssim 0.2$ eV quoted by the Planck Collaboration in Ref.~\cite{Planck:2018vyg} to the tightest bounds $\Sigma \lesssim 0.09$ eV~\cite{Palanque-Delabrouille:2019iyz, DiValentino:2021hoh, Brieden:2022lsd} resulting from joint analyses of Cosmic Microwave Background (CMB) data and late-time expansion history probes in the form of distance measurements from Baryon Acoustic Oscillations (BAO) and Type Ia Supernovae (SN). While the most constraining results placed the IO under some tension with the cosmological upper limits on $\Sigma$ (which were nearing the lower limit of $\sim 0.06$ eV set by oscillations for NO), the exact preference for the NO over the IO (and more broadly the tightest limits themselves) varied significantly depending on the number of underlying assumptions and datasets involved in the different analyses. 

In April 2024, the Dark Energy Spectroscopic Instrument (DESI) released BAO measurements based on its first year of observations~\cite{DESI:2024mwx, DESI:2024uvr, DESI:2024kob, DESI:2024lzq, DESI:2024aqx}. DESI BAO provides precise constraints on the transverse comoving distance, the Hubble rate, and their combination (all relative to the value of the sound horizon at the drag epoch) that, when combined with CMB data from the Planck satellite and the Atacama Cosmology Telescope (ACT)~\cite{ACT:2020gnv,ACT:2023kun,ACT:2023dou}, generally push the best fit to $\Sigma = 0$ and set a particularly tight limit $\Sigma < 0.072\,{\text{eV}}$ at 95\% CL. This limit is well below the minimum allowed by oscillation experiments for IO and close to the minimum for NO~\cite{Jiang:2024viw}. Even more puzzling is that when relaxing the well-motivated physical prior $\Sigma > 0$ (employed in the DESI analyses) and allowing for the unphysical region $\Sigma < 0$, negative neutrino masses appear favored by the data, as reported by a few independent groups~\cite{Craig:2024tky,Naredo-Tuero:2024sgf,Green:2024xbb,Elbers:2024sha}.

Taken at face value, these results raise important concerns, either pointing to possible undetected observational systematics in the DESI BAO data (and/or other probes involved in the analyses) or suggesting the emergence of new physics beyond the $\Lambda$CDM model, adding to a longstanding series of tensions challenging the validity of the standard cosmological framework. Given the current uncertainty, with analyses supporting each direction~\cite{Herold:2024nvk,Escudero:2024uea,RoyChoudhury:2024wri,DESI:2024hhd,Reboucas:2024smm,Elbers:2024sha,Upadhye:2024ypg,Shao:2024mag,Naredo-Tuero:2024sgf,Craig:2024tky,Allali:2024aiv,Green:2024xbb,DESI:2024mwx}, both possibilities warrant serious consideration.



Regarding observational systematics,  a first particularly notable issue arises from the DESI BAO measurement at $z = 0.71$, which exhibits a $\sim3\sigma$ tension with predictions from the Planck best-fit $\Lambda$CDM cosmology~\cite{Wang:2024pui,Colgain:2024xqj,Naredo-Tuero:2024sgf,Sapone:2024ltl}. More broadly, the overall DESI BAO best-fit $\Lambda$CDM cosmology is found to be in $\sim 1.8\sigma$ disagreement with the best-fit Planck $\Lambda$CDM cosmology. This has led many to argue that the observed preference for small $\Sigma$ may stem from a combination of measurements that do not fully align yet are not in significant tension either, representing a classic edge-case scenario. Similarly, systematic uncertainties in Planck CMB data could also be crucial. The DESI collaboration's analysis relies on the \texttt{Plik} likelihood for high-$\ell$ temperature and polarization anisotropies, used in Planck 2018 Data Release 3 (PR3). Over time, mild anomalies have been identified in PR3 data, possibly affecting the constraints on the total neutrino mass. One notable example is the higher lensing amplitude inferred from Planck temperature and polarization spectra, characterized by the parameter $A_{\rm lens}$ \cite{Calabrese:2008rt}, which deviates from its expected value by about $2.8\sigma$ \cite{Planck:2018vyg,DiValentino:2015bja,Renzi:2017cbg,Domenech:2020qay}. Since massive neutrinos impact the CMB spectra similarly to variations in $A_{\rm lens}$, this anomaly has significant implications for neutrino mass constraints \cite{Capozzi:2021fjo}.  To address these concerns, substantial reanalyses of Planck data have been conducted since 2018. The 2020 Planck Data Release 4 (PR4) introduced the \texttt{NPIPE} CMB maps, incorporating improvements that led to updated likelihoods for temperature and polarization spectra. The latest \texttt{CamSpec} \cite{Rosenberg:2022sdy} and \texttt{HiLLiPoP} \cite{Tristram:2023haj} likelihoods, now based on PR4 \texttt{NPIPE} maps, reduce small-scale noise relative to \texttt{Plik} and enhance cosmological parameter constraints by up to 10\%. Both likelihoods indicate a shift toward $A_{\rm lens} = 1$, mitigating the lensing anomaly.  Several groups have explored how the DESI collaboration's neutrino mass bounds, derived using the \texttt{Plik} PR3 likelihood, change when adopting PR4-based likelihoods. These studies suggest that using \texttt{CamSpec} or \texttt{HiLLiPoP} relaxes the neutrino mass constraints compared to \texttt{Plik}~\cite{Allali:2024aiv} while mitigating the shift toward a best-fit $\Sigma = 0$ cosmology~\cite{Naredo-Tuero:2024sgf}.
 
As for new physics, cosmological constraints on $\Sigma$ are inherently model-dependent. While most upper limits on $\Sigma$ are derived within the minimal 7-parameter $\Lambda$CDM+$\Sigma$ model, the tensions emerging in the neutrino sector after the release of the DESI BAO data may indicate limitations of the standard $\Lambda$CDM cosmology in accurately describing these precise observations. The strongest hint of new physics emerging from combining DESI BAO data with other probes involves the Dark Energy (DE) sector. In particular, the DESI collaboration, along with independent re-analyses, reported moderate to strong evidence for Dynamical Dark Energy (DDE) when combining DESI BAO with Planck CMB and Supernovae (SN) distance modulus measurements from three samples: Pantheon-plus \cite{Scolnic:2021amr, Brout:2022vxf}, Union3 \cite{Rubin:2023ovl}, and DESy5 \cite{DES:2024tys, DES:2024upw, DES:2024hip}. Under the Chevallier-Polarski-Linder (CPL) parameterization, $w(a) = w_0 + w_a (1 - a)$, these combinations of datasets consistently point toward a present-day quintessence-like equation of state ($w_0 > -1$) that crosses the phantom barrier ($w_a < 0$). The evidence for DDE reaches $\sim 3.9\sigma$ combining Planck CMB, DESI BAO, and DESy5 SN, while it decreases to $\sim 2.5\sigma$ ($\sim 3.5\sigma$) when replacing DESy5 with Pantheon-plus (Union3) \cite{DESI:2024mwx}. Further studies show that in a 9-parameter $w_0w_a$CDM+$\Sigma$ model, the bound on $\Sigma$ is relaxed compared to the $\Lambda$CDM+$\Sigma$ model \cite{DESI:2024mwx, Elbers:2024sha, Jiang:2024viw}. This emphasizes that it is premature to question our understanding of neutrino physics or make definitive claims about mass ordering based on current results. It also highlights that potential models of new physics that reconcile cosmology with oscillation experiments involve adjustments to cosmological scenarios without altering the standard $3\nu$ paradigm. At the same time, these findings make it imperative to thoroughly investigate the potential implications of these hints of new physics for the global analysis conducted in this study.


\begin{table}[t]
\resizebox{.85\textwidth}{!}{\begin{minipage}{.95\textwidth}
\caption{\label{Tab:Cosmo} 
\footnotesize Results of the cosmological data analysis under three model assumptions: standard cosmology with neutrino masses ($\Lambda$CDM+$\Sigma$), an extended model accounting for lensing systematics ($\Lambda$CDM+$\Sigma$+$A_\mathrm{lens}$), and a nonstandard cosmology with dynamical dark energy and neutrino masses ($w_0w_a$CDM+$\Sigma$). The datasets used are listed in Section~\ref{Sec:Nonosc3}. For Planck, we consider both Plik and CamSpec likelihoods, which yield very similar results in all cases (shown explicitly only for $\Lambda$CDM+$\Sigma$). Upper bounds on $\Sigma$ are reported at the $2\sigma$ level.}
\vspace*{0mm}
\centering
\begin{ruledtabular}
\begin{tabular}{clll}
\# & Model & Data set & $\Sigma$ ($2\sigma$)   \\[1mm]
\hline
 1 & $\Lambda\mathrm{CDM}+\Sigma$				& Plik 									& $<0.175$ eV \\
 2 &											& Plik+DESI								& $<0.065$ eV \\
 3 &											& Plik+DESI+PP							& $<0.073$ eV \\
 4 &											& Plik+DESI+DESy5						& $<0.091$ eV \\
 5 &											& camspec 								& $<0.193$ eV \\
 6 &											& camspec+DESI							& $<0.064$ eV \\
 7 &											& camspec+DESI+PP						& $<0.074$ eV \\
 8 &											& camspec+DESI+DESy5					& $<0.088$ eV \\
\hline
 9 & $\Lambda$CDM+$\Sigma$+$A_\mathrm{lens}$ 	& Plik 									& $<0.616$ eV \\					
 10 &											& Plik+DESI								& $<0.204$ eV \\ 
 11 &											& Plik+DESI+PP							& $<0.255$ eV \\
 12 &											& Plik+DESI+DESy5						& $<0.287$ eV \\
\hline
 13 & $w_0w_a$CDM+$\Sigma$ 						& Plik 									& $<0.279$ eV \\					
 14 &											& Plik+DESI								& $<0.211$ eV \\ 
 15 &											& Plik+DESI+PP							& $<0.155$ eV \\
 16 &											& Plik+DESI+DESy5						& $<0.183$ eV \\
\end{tabular}
\end{ruledtabular}
\end{minipage}}
\end{table}

Table~\ref{Tab:Cosmo} provides updated cosmological constraints on $\Sigma$. These constraints are derived within different cosmological models and using various combinations of cosmological and astrophysical datasets/likelihoods, which have been strategically chosen to account for several possible sources of uncertainty, including potential undetected observational systematics and signals of new physics discussed thus far. Model-wise we consider three different scenarios:

\begin{itemize}
\item $\mathbf{\Lambda}$\textbf{CDM}+$\mathbf{\Sigma}$: We start from the 7-parameter model, which extends the baseline $\Lambda$CDM cosmology by including the total neutrino mass $\Sigma$ as the only additional free parameter. For simplicity and to ensure numerical convergence in the analyses, we adopt the degenerate neutrino mass approximation for $\Sigma$ in this and the other models discussed below. While in the previous global analysis we considered the split masses for NO and IO~\cite{Capozzi:2021fjo}, we anticipate that the variations in the constraints on $\Sigma$ discussed here are significantly larger than the uncertainties introduced by the degenerate mass assumption. This has been repeatedly verified in the literature, including in recent studies by some of us in relation to the latest data~\cite{Jiang:2024viw}.
\item $\mathbf{\Lambda}$\textbf{CDM}+$\mathbf{\Sigma}$+$\mathbf{A_\mathrm{lens}}$: We consider the 8-parameter model, which extends the baseline $\Lambda$CDM cosmology by including the total neutrino mass $\Sigma$ and the lensing amplitude parameter $A_{\rm lens}$. 
\item $\mathbf{w_0w_a}$\textbf{CDM}+$\mathbf{\Sigma}$: We consider the 9-parameter model, which extends the baseline $\Lambda$CDM cosmology by including the total neutrino mass $\Sigma$ and promoting the DE equation of state to a dynamical function of the scale factor $w(a) = w_0 + w_a (1-a)$ characterized by two additional parameters $w_0$ and $w_a$.  
\end{itemize} 

\noindent Data-wise our baseline datasets involve:

\begin{itemize}
\item \textbf{Plik:} The \texttt{Plik} likelihood for high-$\ell$ TT, TE, EE PR3 spectra, along with the \texttt{Commander} and \texttt{SimAll} likelihoods for the low-$\ell$ temperature and polarization TT and EE spectra from the same data release~\cite{Planck:2018vyg}, and a combination of Planck PR4 \texttt{NPIPE} and \texttt{ACT-DR6} lensing likelihoods~~\cite{Carron:2022eyg,ACT:2023kun,ACT:2023dou}. 
\item \textbf{Camspec:} We replace the Planck-PR3 \texttt{Plik} high-$\ell$ TTTEEE likelihood with the more recent \texttt{CamSpec} likelihood~\cite{Rosenberg:2022sdy}, based on the Planck-PR4 \texttt{NPIPE} data release~\cite{Planck:2020olo,Carron:2022eyg}. We retain the same low-$\ell$ and lensing likelihoods from the \textbf{Plik} combination.  
\item \textbf{DESI:} BAO measurements extracted from observations of galaxies and quasars~\cite{DESI:2024uvr}, along with Lyman-$\alpha$~\cite{DESI:2024lzq} tracers from the first year of observations using the Dark Energy Spectroscopic Instrument. These include measurements of the transverse comoving distance, the Hubble horizon, and the angle-averaged distance, as summarized in Tab.~I of Ref.~\cite{DESI:2024mwx}.
\item \textbf{PantheonPlus:} Distance modulus measurements of 1701 light curves from 1550 spectroscopically confirmed Type Ia SN, sourced from eighteen different surveys, gathered from the Pantheon-plus sample~\cite{Scolnic:2021amr, Brout:2022vxf}.
\item \textbf{DESy5:} Distance modulus measurements of 1635 Type Ia SN covering the redshift range $0.10 < z < 1.13$, collected during the full five years of the Dark Energy Survey (DES) Supernova Program~\cite{DES:2024tys,DES:2024upw,DES:2024hip}, along with 194 low-redshift SN in the redshift range $0.025 < z < 0.1$, in common with Pantheon-plus.
\end{itemize} 

Note that we consider different data combinations to assess their impact on constraints on $\Sigma$. We first examine the case where only CMB observations are included, using either the combination of datasets Plik or Camspec. This provides us with a baseline reference case to evaluate how distance measurements from BAO and SN affect the constraints on $\Sigma$. The second combination we analyze is CMB with DESI BAO, which generally yields the tightest limits on $\Sigma$. Finally, we consider the full combination of CMB, DESI BAO, and SN. The inclusion of SN data is particularly relevant, as much of the preference for DDE is driven by SN measurements and reinforced by DESI BAO data \cite{Efstathiou:2024xcq,Notari:2024zmi,Giare:2025pzu}. This makes SN data crucial for evaluating how neutrino mass constraints evolve in the presence of a preference for DDE. We focus on the two Type Ia SN samples, PantheonPlus and DESy5, which exhibit the smallest and largest deviations from $\Lambda$CDM, respectively. We do not consider the Union3 sample, as it falls between these two.

To derive cosmological constraints for all the different models and data combinations, we use Markov Chain Monte Carlo (MCMC) analyses, computing the theoretical predictions with \texttt{CAMB} \cite{Lewis:1999bs, Howlett:2012mh}, and exploring the parameter space with the sampler \texttt{Cobaya} \cite{Lewis:2002ah, Lewis:2013hha}. The output from the MCMC analysis (i.e., the posterior probability distribution functions of $\Sigma$) is converted into equivalent $\chi^2$ values. We then quote the corresponding upper bound at $\Delta \chi^2 = 4$ ($2\sigma$) from the best fit, which is either null or compatible with null values within $<1\sigma$ in all cases considered. Note that our $\chi^2$ translation yields upper bounds on $\Sigma$ that differ slightly (typically by less than 10\%) from those obtained directly from the MCMC posterior distributions. This explains the minor differences between our results summarized in Table~\ref{Tab:Cosmo} and those reported in the literature for the same models and data combinations. Despite these small differences, our findings align with expectations based on previous discussions and existing literature.

Focusing on the minimal $\Lambda$CDM+$\Sigma$ model, we find that the Planck PR3-based Plik likelihood leads to an upper bound of $\Sigma < 0.175$ eV. This limit is slightly relaxed to $\Sigma < 0.193$ eV when using Camspec. Although, for clarity, we show the Plik versus Camspec comparison only for the standard $\Lambda$CDM+$\Sigma$ scenario in Table~\ref{Tab:Cosmo}, we have verified that the same effect is consistent across all examined models. Among the different data combinations, the tightest constraint, $\Sigma < 0.065$ eV, comes from combining Planck CMB and DESI BAO. Conversely, introducing SN measurements (whether PantheonPlus or DESy5) leads to weaker constraints. This is due to the fact that, assuming a $\Lambda$CDM cosmology, SN data typically favor a higher present-day matter density, $\Omega_m$, than BAO measurements \cite{Allali:2024aiv,Colgain:2024mtg}. Consequently, when SN data is included, $\Omega_m$ shifts to larger values, which allows for a slightly higher total neutrino mass.

Within the $\Lambda$CDM+$A_{\rm lens}$+$\Sigma$ model, CMB-only results relax the upper bound to $\Sigma < 0.6$ eV. This is approximately a factor of $\sim 3.5$ larger than in the baseline case, underscoring once again the strong correlation between $\Sigma$ and $A_{\rm lens}$. The inclusion of DESI BAO tightens this limit to $\Sigma < 0.204$ eV, an improvement by a factor of $\sim 3$, which is comparable to the relative improvement observed within the baseline $\Lambda$CDM+$\Sigma$ scenario. Similarly to that case, incorporating either PantheonPlus or DESy5 slightly weakens the bound on $\Sigma$ for the same reasons outlined earlier.

In the dynamical $w_0w_a$CDM+$\Sigma$ model, where the cosmological constant assumption is relaxed, CMB data alone yield a constraint of $\Sigma < 0.279$ eV, which is about 1.6 times larger than in the $\Lambda$CDM+$\Sigma$ model. In this case, the inclusion of DESI BAO only marginally improves the bound, as BAO data primarily constrains the $w_0$ and $w_a$ parameters, which significantly affect background expansion but have minimal impact on CMB physics. Notably, incorporating SN measurements does not relax the bound on $\Sigma$ as in other models but instead slightly strengthens it: both PantheonPlus and DESy5 push the limit below 0.2 eV. This occurs because SN data favor a deviation from $\Lambda$CDM, redistributing the preference for higher $\Omega_m$ and tightening the constraint on $\Sigma$.

In conclusion, it appears rather premature to quote a ``consensus'' upper bound on $\Sigma$ from cosmological data at present. We prefer to quote a ``range'' of upper bounds, noticing that the $2\sigma$ cosmological limits on $\Sigma$ from Table~\ref{Tab:Cosmo} cluster around a reasonable
``geometric average'' value of $\Sigma < 0.2$ eV, with variations up to a factor of 3 (upwards or downwards),  depending on the specific model and dataset employed. We thus surmise that
\begin{equation}
\label{Boundsigma}
\Sigma < 0.2~\mathrm{eV\ at\ }2\sigma\ \mathrm{(within\ a\ factor\ of\ 3)}\,
\end{equation}
may be considered as a reasonable summary of acceptable and physically motivated current bounds, ranging from very conservative ($\Sigma < 0.60$~eV) to very aggressive ($\Sigma < 0.07$~eV).  This interval of limits is essentially supported by Planck data alone within standard cosmology, as well as by both Planck and DESI in nonstandard cases. In particular, this range also covers representative effects of observational systematics or of new physics, that may affect cosmology without altering the standard $3\nu$ framework adopted herein. We note that recent discussions within the neutrino and cosmology communities also tend to quote ``reasonably safe'' upper limits clustered around 0.2~eV, see e.g.\ \cite{CERNnuWeek}.  
 

\vspace*{-2mm}
\subsection{Comparison of oscillation and nonoscillation bounds on $(m_{\beta},\,m_{\beta\beta},\,\Sigma)$}
\label{Sec:Nonosc4}

\begin{figure}[t!]
\begin{minipage}[c]{0.8\textwidth}
\includegraphics[width=0.6\textwidth]{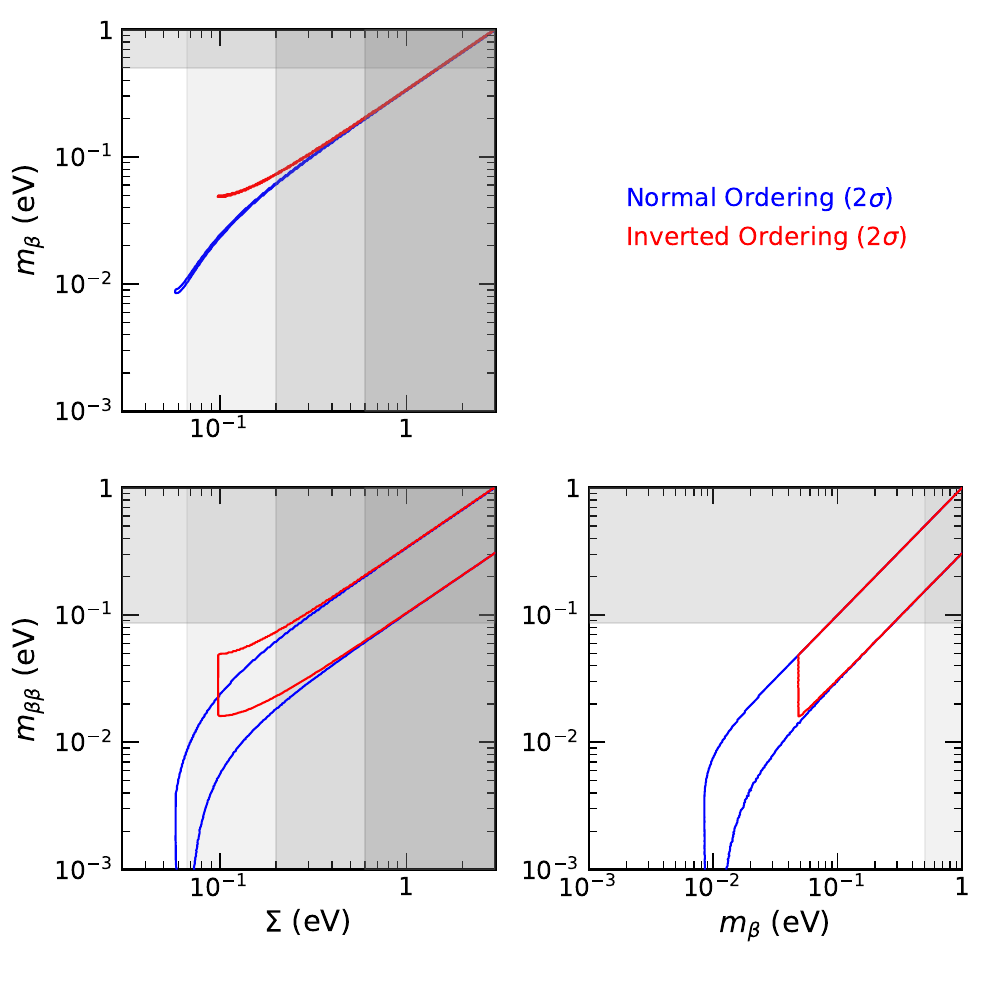}
\vspace*{-4mm}
\caption{\label{Fig_10}
\footnotesize Comparison of separate oscillation and nonoscillation 
bounds on the absolute mass observables $(\Sigma,\,m_\beta,\,m_{\beta\beta})$, in each of the three planes charted by a pair of observables. 
Bounds are shown at $2\sigma$ for NO (blue) and IO (red).  For $\Sigma$
we consider three alternative bounds, corresponding to $\Sigma<0.2$~eV within a factor of 3 (upward and downward),  see the text for details. Note that we take
 $\Delta \chi^2_\mathrm{IO-NO}=0$ in this figure.
}
\end{minipage}
\end{figure}

The constraints derived on the oscillation parameters strongly reduce the phase space available for the three absolute mass observables $(m_\beta,\,m_{\beta\beta},\,\Sigma)$ \cite{Fogli:2004as}. Pairs of observables are then constrained within two (NO and IO) correlated bands, that tend to merge for relatively high neutrino masses as compared to their splittings (so-called ``degenerate'' mass spectrum), while they branch out for relatively small masses (so-called ``hierarchical'' mass spectrum).  The width of the IO and NO bands is determined by the small uncertainty on the oscillation parameters and, for $m_{\beta\beta}$, also by the unknown values of the Majorana phases. In this context, we assume the IO and NO cases on the same footing ($\Delta \chi^2_\mathrm{IO-NO}=0$).

Figure~\ref{Fig_10} shows the $2\sigma$ bands allowed by current oscillation constraints (NO in blue and IO in red) in the planes charted by any pair of observables among $(m_\beta,\,m_{\beta\beta},\,\Sigma)$. Also shown (in grey) are the regions disfavored by the upper limits at $2\sigma$ from   
Eqs.~(\ref{Boundmb}), (\ref{Boundmbb}) and (\ref{Boundsigma}). For the latter we distinguish three limits, corresponding to $\Sigma$ values of 0.2, $0.2\times 3$ and $0.2/3$~eV. In the $(m_\beta,\,\Sigma)$ plane, the constraint $m_{\beta}<0.5$~eV represents a safe and uncontroversial limit that, however, is still relatively weak, as it cuts only the upper part of the NO and IO bands; a stronger cut is placed by the conservative upper limit $\Sigma<0.6$~eV. Similar arguments apply to the $(m_\beta,\,m_{\beta\beta})$ plane, where the bound 
$m_{\beta\beta}<0.085$~eV wins (if neutrinos are Majorana). In any case, all these bounds disfavor a large part of the degenerate spectrum region. 

In the $(m_{\beta\beta},\,\Sigma)$ panel of Fig.~\ref{Fig_10}, the intermediate cosmology bound $\Sigma<0.2$~eV and the $0\nu\beta\beta$ bound $m_{\beta\beta}<0.085$~eV cut the parameter space in the transition region between degenerate and hierarchical, implying individual neutrino masses definitely lighter than one tenth of eV, and leaving open both options for mass ordering (NO and IO). We think that this may be considered as a reasonable summary of the current situation about absolute neutrino masses. This situation is similar to (and even a bit more conservative than) the case dubbed as ``default'' in the previous analysis \cite{Capozzi:2021fjo}: indeed, despite the experimental progress and results on $(m_\beta,\,m_{\beta\beta},\,\Sigma)$ appeared after the work  \cite{Capozzi:2021fjo}, there is currently a more pronounced uncertainty on the dominant limits placed by $\Sigma$. Finally, considering the aggressive upper limit $\Sigma<0.067$~eV in Fig.~\ref{Fig_10}, cosmology alone would not only cut a large part of the hierarchical spectrum region, but also exclude the IO case, leaving only a thin slice of phase space available for NO. In this case, $\beta$ and $0\nu\beta\beta$ decay searches for absolute neutrino masses should be targeted to  values as small as $m_\beta\sim 10^{-2}$~eV and $m_{\beta\beta}<10^{-2}$~eV, respectively. 
A future determination of the mass ordering via oscillations will be important to reduce the variety of possible cases in Fig.~\ref{Fig_10}, but might also reveal  tensions with non-oscillation observables (e.g., lack of overall convergence on either NO or IO) that could even point towards new physics beyond the standard $3\nu$ framework, see e.g.~\cite{Neutel2023}.

\vspace*{-2mm}
\section{Summary and perspectives}
\label{Sec:End}

We have performed an updated global analysis of the known and unknown parameters of the standard $3\nu$ framework, using oscillation and nonoscillation data available at the beginning of 2025. Upper and lower bounds on the known 
mixing angles $(\theta_{12},\,\theta_{23},\,\theta_{13})$ and squared mass gaps $(\delta m^2,\,|\Delta m^2|)$ governing $\nu$ oscillations are summarized in Fig.~\ref{Fig_03} and in Table~\ref{Tab:Synopsis}  for the cases of NO and IO. With respect to the previous 2021 analysis in \cite{Capozzi:2021fjo}, 
the combination of accelerator, reactor and atmospheric $\nu$ data lead to appreciably reduced uncertainties for $\theta_{23}$, $\theta_{13}$ and $|\Delta m^2|$. In particular, $|\Delta m^2|$ is the first $3\nu$ parameter to enter the domain of subpercent precision (0.8\% at $1\sigma$). At this level of accuracy,
issues related to neutrino energy reconstruction, interaction models and correlated systematic uncertainties in accelerator and atmospheric neutrino experiments (and in related data analyses) warrant further joint studies by the collaborations themselves, possibly leading to more conservative error estimates 
as argued in Sec.~\ref{Osc1}. Another motivation for these studies is provided by the expected synergy between JUNO and rest-of-the-world oscillation data in determining the mass ordering via independent $|\Delta m^2|$ measurements with subpercent precision. 
In this context, we have mapped current fit results in the plane charted by $\delta m^2$ 
and by the effective parameter $\Delta m^2_{ee}$ (Fig.~\ref{Fig_07}) to be directly measured by JUNO, with an immediate impact on the NO vs IO issue (as qualitatively discussed in Sec.~\ref{Osc3}). 

The mass ordering, together with the $\theta_{23}$ octant and the CP-violating phase $\delta$, form the set of oscillation unknowns. Current
hints about such unknowns are all slightly weaker than before \cite{Capozzi:2021fjo}, and amount to a preference for NO versus IO at  $2.2\sigma$, for CP violation versus conservation in NO (1.3$\sigma$) and for the first $\theta_{23}$ octant versus the second in NO ($1.1\sigma$). Correlations between 
such results in increasingly rich data sets have been discussed through Figs.~\ref{Fig_04}-\ref{Fig_06}. The current situation about oscillation unknowns 
appears statistically unstable and quite open to different options in the future. We have also discussed the unknown absolute mass observables, including the effective mass $m_\beta$ from $\beta$-decay, the Majorana mass $m_{\beta\beta}$ from $0\nu\beta\beta$ decay, and the total $\nu$ mass $\Sigma$ in cosmology. At a C.L.\ of $2\sigma$ we report the bounds $m_\beta<0.50$~eV and $m_{\beta\beta}<0.086$~eV, where the latter includes parametrized 
nuclear matrix element (co)variances for different isotopes. Concerning $\Sigma$, an updated
discussion of upper bounds necessarily involves a survey of the emerging tensions within the standard $\Lambda$CDM cosmological model, suggesting possible observational systematics or model extensions. We have discussed representative combinations of data, with or without augmenting the $\Lambda$CDM model with extra parameters accounting for possible systematics (lensing anomaly) or new physics (dynamical dark energy). The resulting $2\sigma$  upper limits 
(Table~\ref{Tab:Cosmo}) are roughly spread around the bound $\Sigma < 0.2$~eV within a factor of three (both upwards and downwards), with different implications for NO and IO scenarios, as discussed in the context of Fig.~\ref{Fig_10}. Given such large uncertainties, it is premature to endorse definite conclusions about absolute neutrino masses, although it is reasonable to assume that their individual values are below one tenth of eV --- not much larger than their splittings, 
and compatible with both NO and IO.

We are at a very interesting junction in the development of the $3\nu$ paradigm. On the one hand, the five known observables 
$(\theta_{12},\,\theta_{23},\,\theta_{13})$ and $(\delta m^2,\,|\Delta m^2|)$ have been measured with an accuracy ranging from the few-percent level of $\theta_{23}$ to
the subpercent level of $|\Delta m^2|$. On the other hand, the status of the five $3\nu$ unknowns (mass ordering, $\theta_{23}$ octant, CP violation, absolute masses, Dirac or Majorana nature) remains quite open to different outcomes. The next relevant ``trait d'union'' between knowns and unknowns will be provided by the first JUNO data release, that will not only improve the precision of $\Delta m^2$ and other parameters, but also affect the relative likelihood of NO and IO, 
first in synergy with other data and then autonomously. On a different time scale, 
it will also be important to clarify the situation of cosmological bounds on $\Sigma$, 
especially  in view of a possible detection or claim of $\Sigma>0$ in future  experiments. Of course,
also searching for the unknown CP-violating phase $\delta$ remains a high priority. Along this path,
it should not be given for granted that 
synergies will clearly emerge, or that different datasets will necessarily converge: 
(new) tensions might affect the comparison of independent data, and redundant information from
different and complementary experiments will remain crucial to either confirm the $3\nu$ paradigm, or to find new phenomena beyond it.

\acknowledgments

The work of F.C., E.L, A.M., A.M.\ and A.P.\ was partially supported by the research grant number 2022E2J4RK ``PANTHEON: Perspectives in Astroparticle and Neutrino THEory with Old and New messengers'' under the program PRIN 2022 funded by the Italian Ministero dell'Universit\`a e della Ricerca (MUR) and by the European Union -- Next Generation EU, as well as by the Theoretical Astroparticle
Physics (TAsP) initiative of the Istituto Nazionale di Fisica Nucleare (INFN). 
W.G.\ acknowledges support from the Lancaster-Sheffield Consortium for Fundamental Physics through the Science and Technology Facilities Council (STFC) grant ST/X000621/1. W.G.\ acknowledges IT Services at The University of Sheffield for the provision of services for High Performance Computing.
We thank Eleonora Di Valentino for useful discussions during the preparation of this work.


{}


\begin{thebibliography}{999}

\bibitem{ParticleDataGroup:2024cfk}
S.~Navas \textit{et al.} [Particle Data Group],
``Review of Particle Physics,''
Phys. Rev. D \textbf{110}, no.3, 030001 (2024)




\bibitem{PDG1}
M.~C.\ Gonzalez-Garcia and M.\ Yokoyama, ``Neutrino Masses, Mixing, and Oscillations,'' 
in \cite{ParticleDataGroup:2024cfk}.

\bibitem{Nu2024}
{\em Neutrino 2024}, XXXI International Conference on Neutrino Physics and Astrophysics (Milan, Italy, 16--22 June 2024). Website:
https://agenda.infn.it/event/37867/

\bibitem{McDonald24} A.~McDonald, ``Opening talk: Where are we?,'' in \cite{Nu2024}.

\bibitem{Giganti24} C.~Giganti [T2K], ``T2K recent results and plans,'' in \cite{Nu2024}.

\bibitem{Wolcott24} J.~Wolcott [NOvA], ``New NOvA Results with 10 Years of Data,'' in \cite{Nu2024}.

\bibitem{Satellite} {\em Multi-experiment oscillation measurements}, NuFACT Satellite Workshop 
(Argonne Nat.\ Lab., US, 15-Sept.~2024). Website:  https://indico.global/event/3065/

\bibitem{Fogli:1993ck}
G.~L.~Fogli, E.~Lisi and D.~Montanino,
``A comprehensive analysis of solar, atmospheric, accelerator and reactor neutrino experiments in a hierarchical three generation scheme,''
Phys. Rev. D \textbf{49}, 3626-3642 (1994).

\bibitem{Tortola24} M.~Tortola, ``Global analysis of three-neutrino oscillations,'' in \cite{Nu2024}.

\bibitem{Lisi24} E.~Lisi,  ``Outlook on $\nu$ theory and phenomenology,'' in \cite{Nu2024}.

\bibitem{Schwetz24}
T.~Schwetz, ``Global analyses now and in the future,'' Concluding Talk at {\em NOW 2024}, 
International Neutrino Oscillation Workshop (Otranto, Lecce, Italy, 2--8 Sept.~2024). Website:
https://agenda.infn.it/event/39753/

\bibitem{deSalas:2020pgw}
P.~F.~de Salas, D.~V.~Forero, S.~Gariazzo, P.~Mart\'\i{}nez-Mirav\'e, O.~Mena, C.~A.~Ternes, M.~T\'ortola and J.~W.~F.~Valle,
``2020 global reassessment of the neutrino oscillation picture,''
JHEP \textbf{02}, 071 (2021)
[arXiv:2006.11237 [hep-ph]].

\bibitem{Esteban:2020cvm}
I.~Esteban, M.~C.~Gonzalez-Garcia, M.~Maltoni, T.~Schwetz and A.~Zhou,
``The fate of hints: updated global analysis of three-flavor neutrino oscillations,''
JHEP \textbf{09}, 178 (2020)
[arXiv:2007.14792 [hep-ph]].


\bibitem{Capozzi:2021fjo}
F.~Capozzi, E.~Di Valentino, E.~Lisi, A.~Marrone, A.~Melchiorri and A.~Palazzo,
``Unfinished fabric of the three neutrino paradigm,''
Phys. Rev. D \textbf{104}, no.8, 083031 (2021)
[arXiv:2107.00532 [hep-ph]].

\bibitem{Esteban:2024eli}
I.~Esteban, M.~C.~Gonzalez-Garcia, M.~Maltoni, I.~Martinez-Soler, J.~P.~Pinheiro and T.~Schwetz,
``NuFit-6.0: updated global analysis of three-flavor neutrino oscillations,''
JHEP \textbf{12}, 216 (2024)
[arXiv:2410.05380 [hep-ph]].

\bibitem{Cao24} J.~Cao, ``Status of JUNO,'' in \cite{Nu2024}


\bibitem{JUNO:2015zny}
F.~An \textit{et al.} [JUNO],
``Neutrino Physics with JUNO,''
J. Phys. G \textbf{43}, no.3, 030401 (2016)
[arXiv:1507.05613 [physics.ins-det]].

\bibitem{JUNO:2022mxj}
A.~Abusleme \textit{et al.} [JUNO],
``Sub-percent precision measurement of neutrino oscillation parameters with JUNO,''
Chin. Phys. C \textbf{46}, no.12, 123001 (2022)
[arXiv:2204.13249 [hep-ex]].

\bibitem{JUNO:2024jaw}
A.~Abusleme \textit{et al.} [JUNO],
``Potential to Identify the Neutrino Mass Ordering with Reactor Antineutrinos in JUNO,''
[arXiv:2405.18008 [hep-ex]].


\bibitem{Elbers24} W.~Elbers, ``First neutrino results from the Dark Energy Spectroscopic Instrument (DESI),'' in \cite{Nu2024}.

\bibitem{Gariazzo24} S.~Gariazzo,  ``Neutrino nonstandard scenarios and cosmology,'' in \cite{Nu2024}.

\bibitem{Jiang:2024viw}
J.~Q.~Jiang, W.~Giar\`e, S.~Gariazzo, M.~G.~Dainotti, E.~Di Valentino, O.~Mena, D.~Pedrotti, S.~S.~da Costa and S.~Vagnozzi,
``Neutrino cosmology after DESI: tightest mass upper limits, preference for the normal ordering, and tension with terrestrial observations,''
[arXiv:2407.18047 [astro-ph.CO]].


\bibitem{Menendez24}
J.\ Menendez, ``Neutrinoless $\beta\beta$ decay searches: theory of nuclear matrix elements,'' in \cite{Nu2024}.


\bibitem{Fogli:2012ua}
G.~L.~Fogli, E.~Lisi, A.~Marrone, D.~Montanino, A.~Palazzo and A.~M.~Rotunno,
``Global analysis of neutrino masses, mixings and phases: entering the era of leptonic CP violation searches,''
Phys. Rev. D \textbf{86}, 013012 (2012)
[arXiv:1205.5254 [hep-ph]].


\bibitem{Super-Kamiokande:2023jbt}
K.~Abe \textit{et al.} [Super-Kamiokande],
``Solar neutrino measurements using the full data period of Super-Kamiokande-IV,''
Phys. Rev. D \textbf{109}, no.9, 092001 (2024)
[arXiv:2312.12907 [hep-ex]].

\bibitem{SSM2023}
Y.~Herrera and A.~Serenelli, ``Standard Solar Models B23/SF-III'' (2023), available at the website: https://doi.org/10.5281/zenodo.10174170 

\bibitem{MB22m}  We adopt the solar model denoted as ``MB22m'' in \cite{SSM2023}, based on the solar meteoritic composition in: 
E.~Magg \textit{et al.}, ``Observational constraints on the origin of the elements.\ IV.\ Standard composition of the Sun,''
Astronomy \& Astrophysics \textbf{661}, A140 (2022)
[arXiv:2203.02255 [astro-ph]]

\bibitem{Haxton:2025hye}
W.~C.~Haxton and E.~Rule,
``The gallium solar neutrino capture cross section revisited,''
Phys. Lett. B \textbf{861}, 139259 (2025)
[arXiv:2501.03528 [nucl-ex]].


\bibitem{T2K:2023smv}
K.~Abe \textit{et al.} [T2K],
``Measurements of neutrino oscillation parameters from the T2K experiment using \ensuremath{3.6\times 10^{21}} protons on target,''
Eur. Phys. J. C \textbf{83}, no.9, 782 (2023)
[arXiv:2303.03222 [hep-ex]].

\bibitem{T2K:2023mcm}
K.~Abe \textit{et al.} [T2K],
``Updated T2K measurements of muon neutrino and antineutrino disappearance using \ensuremath{3.6\times 10^{21}} protons on target,''
Phys. Rev. D \textbf{108}, no.7, 072011 (2023)
[arXiv:2305.09916 [hep-ex]].

\bibitem{Jargowsky:2024mwc}
For NOvA results, see also B.~J.~Jargowsky,
``A Measurement of \ensuremath{\nu_e} Appearance and \ensuremath{\nu_\mu} Disappearance Using 10 Years of Data from the NOvA Experiment,''
PhD Thesis (U.\ of California, Irvine, 2024), available at https://escholarship.org/uc/item/8t19p34n


\bibitem{DayaBay:2022orm}
F.~P.~An \textit{et al.} [Daya Bay],
``Precision Measurement of Reactor Antineutrino Oscillation at Kilometer-Scale Baselines by Daya Bay,''
Phys. Rev. Lett. \textbf{130}, no.16, 161802 (2023)
[arXiv:2211.14988 [hep-ex]]. See also the 
Supplemental Material available at http://link.aps.org/supplemental/10.1103/PhysRevLett.130.161802

\bibitem{RENO:2024msr}
S.~Jeon \textit{et al.} [RENO],
``Measurement of reactor antineutrino oscillation amplitude and frequency using 3800 days of complete data sample of the RENO experiment,''
[arXiv:2412.18711 [hep-ex]].

\bibitem{Fogli:2003th}
G.~L.~Fogli, E.~Lisi, A.~Marrone and D.~Montanino,
``Status of atmospheric $\nu_\mu\to\nu_\tau$ oscillations and decoherence after the first K2K spectral data,''
Phys. Rev. D \textbf{67}, 093006 (2003)
[arXiv:hep-ph/0303064 [hep-ph]].


\bibitem{Capozzi:2018ubv}
F.~Capozzi, E.~Lisi, A.~Marrone and A.~Palazzo,
``Current unknowns in the three neutrino framework,''
Prog. Part. Nucl. Phys. \textbf{102}, 48-72 (2018)
[arXiv:1804.09678 [hep-ph]].

\bibitem{Super-Kamiokande:2023ahc}
T.~Wester \textit{et al.} [Super-Kamiokande],
``Atmospheric neutrino oscillation analysis with neutron tagging and an expanded fiducial volume in Super-Kamiokande I\textendash{}V,''
Phys. Rev. D \textbf{109}, no.7, 072014 (2024)
[arXiv:2311.05105 [hep-ex]].

\bibitem{SKmap}
Super-Kamiokande Collaboration, Data Release for \cite{Super-Kamiokande:2023ahc} (2023), available at https://zenodo.org/records/8401262


\bibitem{IceCube:2024xjj}
R.~Abbasi \textit{et al.} [IceCube],
``Measurement of atmospheric neutrino oscillation parameters using convolutional neural networks with 9.3 years of data in IceCube DeepCore,''
[arXiv:2405.02163 [hep-ex]].


\bibitem{ICmap} IceCube Collaboration, Data Release for \cite{IceCube:2024xjj} (2024), available at https://doi.org/10.7910/DVN/U20MMB

\bibitem{SKatmThesis} T.~Wester, ``Discerning the neutrino mass ordering
using atmospheric neutrinos in Super-Kamiokande I\textendash{}V,'' PhD Thesis (Boston U., 2023), available at 
https://open.bu.edu/handle/2144/46427


\bibitem{ICThesis} M. Prado Rodriguez, ``Measurement of the Neutrino Mass Ordering with 9.28 Years of IceCube DeepCore Data,''
PhD Thesis (U.\ of Wisconsins--Madison, 2024), available at https://digital.library.wisc.edu/1711.dl/24CPK3B2BJYH586

\bibitem{Fogli:2002pt}
G.~L.~Fogli, E.~Lisi, A.~Marrone, D.~Montanino and A.~Palazzo,
``Getting the most from the statistical analysis of solar neutrino oscillations,''
Phys. Rev. D \textbf{66}, 053010 (2002)
[arXiv:hep-ph/0206162 [hep-ph]].

\bibitem{LisiNPB24} E.\ Lisi, ``Global analysis of neutrino mass-mixing parameters: What Next?,'' in
{\em NPB 2024}, International Symposium on Neutrino Physics and Beyond (Hong Kong, China, 19-21 Feb.~2024).
Website: https://indico.ihep.ac.cn/event/20514/

\bibitem{Arguelles:2022hrt}
C.~A.~Arg\"uelles, P.~Fern\'andez, I.~Mart\'\i{}nez-Soler and M.~Jin,
``Measuring Oscillations with a Million Atmospheric Neutrinos,''
Phys. Rev. X \textbf{13}, no.4, 041055 (2023)
[arXiv:2211.02666 [hep-ph]].


\bibitem{T2K:2024wfn}
K.~Abe \textit{et al.} [T2K and Super-Kamiokande],
``First Joint Oscillation Analysis of Super-Kamiokande Atmospheric and T2K Accelerator Neutrino Data,''
Phys. Rev. Lett. \textbf{134}, no.1, 011801 (2025)
[arXiv:2405.12488 [hep-ex]].

\bibitem{Abe:2024avs}
S.~Abe,
``Implementation and investigation of electron-nucleus scattering in the NEUT neutrino event generator,''
Phys. Rev. D \textbf{111}, no.3, 033006 (2025)
[arXiv:2412.07466 [hep-ph]].

\bibitem{Coyle:2025xjk}
N.~M.~Coyle, S.~W.~Li and P.~A.~N.~Machado,
``Neutrino-Nucleus Cross Section Impacts on Neutrino Oscillation Measurements,''
[arXiv:2502.19467 [hep-ph]].


\bibitem{Chatterjee:2024kbn}
S.~S.~Chatterjee and A.~Palazzo,
``Status of tension between NOvA and T2K after Neutrino 2024 and possible role of nonstandard neutrino interactions,''
Phys. Rev. D \textbf{110}, no.11, 113002 (2024)
[arXiv:2409.10599 [hep-ph]].


\bibitem{Fogli:2008jx}
G.~L.~Fogli, E.~Lisi, A.~Marrone, A.~Palazzo and A.~M.~Rotunno,
``Hints of $\theta_{13}$ \ensuremath{>} 0 from global neutrino data analysis,''
Phys. Rev. Lett. \textbf{101}, 141801 (2008)
[arXiv:0806.2649 [hep-ph]].

\bibitem{Fogli:2011qn}
G.~L.~Fogli, E.~Lisi, A.~Marrone, A.~Palazzo and A.~M.~Rotunno,
``Evidence of $\theta_{13}$\ensuremath{>}0 from global neutrino data analysis,''
Phys. Rev. D \textbf{84}, 053007 (2011)
[arXiv:1106.6028 [hep-ph]].

\bibitem{DayaBay:2012fng}
F.~P.~An \textit{et al.} [Daya Bay],
``Observation of electron-antineutrino disappearance at Daya Bay,''
Phys. Rev. Lett. \textbf{108}, 171803 (2012)
[arXiv:1203.1669 [hep-ex]].

\bibitem{RENO:2012mkc}
J.~K.~Ahn \textit{et al.} [RENO],
``Observation of Reactor Electron Antineutrino Disappearance in the RENO Experiment,''
Phys. Rev. Lett. \textbf{108}, 191802 (2012)
[arXiv:1204.0626 [hep-ex]].

\bibitem{DoubleChooz:2012gmf}
Y.~Abe \textit{et al.} [Double Chooz],
``Reactor electron antineutrino disappearance in the Double Chooz experiment,''
Phys. Rev. D \textbf{86}, 052008 (2012)
[arXiv:1207.6632 [hep-ex]].

\bibitem{SNO:2024wzq}
A.~Allega \textit{et al.} [SNO+],
``Initial measurement of reactor antineutrino oscillation at SNO+,''
Eur. Phys. J. C \textbf{85}, no.1, 17 (2025)
[arXiv:2405.19700 [hep-ex]].

\bibitem{Maneira24}
J.~Maneira, ``Solar Neutrinos: Recent Results and Prospects,''
in \cite{Nu2024}.

\bibitem{Petcov:2001sy}
S.~T.~Petcov and M.~Piai,
``The LMA MSW solution of the solar neutrino problem, inverted neutrino mass hierarchy and reactor neutrino experiments,''
Phys. Lett. B \textbf{533}, 94-106 (2002)
[arXiv:hep-ph/0112074 [hep-ph]].

\bibitem{Nunokawa:2005nx}
H.~Nunokawa, S.~J.~Parke and R.~Zukanovich Funchal,
``Another possible way to determine the neutrino mass hierarchy,''
Phys. Rev. D \textbf{72}, 013009 (2005)
[arXiv:hep-ph/0503283 [hep-ph]].


\bibitem{Minakata:2007tn}
H.~Minakata, H.~Nunokawa, S.~J.~Parke and R.~Zukanovich Funchal,
``Determination of the Neutrino Mass Hierarchy via the Phase of the Disappearance Oscillation Probability with a Monochromatic $\bar{\nu}_e$ Source,''
Phys. Rev. D \textbf{76}, 053004 (2007)
[erratum: Phys. Rev. D \textbf{76}, 079901 (2007)]
[arXiv:hep-ph/0701151 [hep-ph]].

\bibitem{Capozzi:2013psa}
F.~Capozzi, E.~Lisi and A.~Marrone,
``Neutrino mass hierarchy and electron neutrino oscillation parameters with one hundred thousand reactor events,''
Phys. Rev. D \textbf{89}, no.1, 013001 (2014)
[arXiv:1309.1638 [hep-ph]].

\bibitem{Capozzi:2015bpa}
F.~Capozzi, E.~Lisi and A.~Marrone,
``Neutrino mass hierarchy and precision physics with medium-baseline reactors: Impact of energy-scale and flux-shape uncertainties,''
Phys. Rev. D \textbf{92}, no.9, 093011 (2015)
[arXiv:1508.01392 [hep-ph]].


\bibitem{Capozzi:2020cxm}
F.~Capozzi, E.~Lisi and A.~Marrone,
``Mapping reactor neutrino spectra from TAO to JUNO,''
Phys. Rev. D \textbf{102}, no.5, 056001 (2020)
[arXiv:2006.01648 [hep-ph]].

\bibitem{deGouvea:2005hk}
A.~de Gouvea, J.~Jenkins and B.~Kayser,
``Neutrino mass hierarchy, vacuum oscillations, and vanishing $|U_{e3}|$,''
Phys. Rev. D \textbf{71}, 113009 (2005)
[arXiv:hep-ph/0503079 [hep-ph]].


\bibitem{Minakata:2006gq}
H.~Minakata, H.~Nunokawa, S.~J.~Parke and R.~Zukanovich Funchal,
``Determining neutrino mass hierarchy by precision measurements in electron and muon neutrino disappearance experiments,''
Phys. Rev. D \textbf{74}, 053008 (2006)
[arXiv:hep-ph/0607284 [hep-ph]].


\bibitem{Qian:2012xh}
X.~Qian, D.~A.~Dwyer, R.~D.~McKeown, P.~Vogel, W.~Wang and C.~Zhang,
``Mass Hierarchy Resolution in Reactor Anti-neutrino Experiments: Parameter Degeneracies and Detector Energy Response,''
Phys. Rev. D \textbf{87}, no.3, 033005 (2013)
[arXiv:1208.1551 [physics.ins-det]].


\bibitem{Li:2013zyd}
Y.~F.~Li, J.~Cao, Y.~Wang and L.~Zhan,
``Unambiguous Determination of the Neutrino Mass Hierarchy Using Reactor Neutrinos,''
Phys. Rev. D \textbf{88}, 013008 (2013)
[arXiv:1303.6733 [hep-ex]].


\bibitem{Cabrera:2020ksc}
A.~Cabrera, Y.~Han, M.~Obolensky, F.~Cavalier, J.~Coelho, D.~Navas-Nicol\'as, H.~Nunokawa, L.~Simard, J.~Bian and N.~Nayak, \textit{et al.}
``Synergies and prospects for early resolution of the neutrino mass ordering,''
Sci. Rep. \textbf{12}, no.1, 5393 (2022)
[arXiv:2008.11280 [hep-ph]].


\bibitem{Cao:2020ans}
S.~Cao, A.~Nath, T.~V.~Ngoc, P.~T.~Quyen, N.~T.~Hong Van and N.~K.~Francis,
``Physics potential of the combined sensitivity of T2K-II, NO$\nu$A extension, and JUNO,''
Phys. Rev. D \textbf{103}, no.11, 112010 (2021)
[arXiv:2009.08585 [hep-ph]].



\bibitem{Choubey:2022gzv}
S.~Choubey, M.~Ghosh and D.~Raikwal,
``Neutrino mass ordering: Circumventing the challenges using synergy between T2HK and JUNO,''
Phys. Rev. D \textbf{106}, no.11, 115013 (2022)
[arXiv:2207.04784 [hep-ph]].

\bibitem{Blennow:2013vta}
M.~Blennow and T.~Schwetz,
``Determination of the neutrino mass ordering by combining PINGU and Daya Bay II,''
JHEP \textbf{09}, 089 (2013)
[arXiv:1306.3988 [hep-ph]].

\bibitem{IceCube-Gen2:2019fet}
M.~G.~Aartsen \textit{et al.} [IceCube-Gen2],
``Combined sensitivity to the neutrino mass ordering with JUNO, the IceCube Upgrade, and PINGU,''
Phys. Rev. D \textbf{101}, no.3, 032006 (2020)
[arXiv:1911.06745 [hep-ex]].


\bibitem{KM3NeT:2021rkn}
S.~Aiello \textit{et al.} [KM3NeT and JUNO],
``Combined sensitivity of JUNO and KM3NeT/ORCA to the neutrino mass ordering,''
JHEP \textbf{03}, 055 (2022)
[arXiv:2108.06293 [hep-ex]].


\bibitem{Ghosh:2012px}
A.~Ghosh, T.~Thakore and S.~Choubey,
``Determining the Neutrino Mass Hierarchy with INO, T2K, NOvA and Reactor Experiments,''
JHEP \textbf{04}, 009 (2013)
[arXiv:1212.1305 [hep-ph]].

\bibitem{Raikwal:2022nqk}
D.~Raikwal, S.~Choubey and M.~Ghosh,
``Determining neutrino mass ordering with ICAL, JUNO and T2HK,''
Eur. Phys. J. Plus \textbf{138}, no.2, 110 (2023)
[erratum: Eur. Phys. J. Plus \textbf{138}, no.6, 485 (2023)]
[arXiv:2207.06798 [hep-ph]].

\bibitem{Qian:2015waa}
X.~Qian and P.~Vogel,
``Neutrino Mass Hierarchy,''
Prog. Part. Nucl. Phys. \textbf{83}, 1-30 (2015)
[arXiv:1505.01891 [hep-ex]].



\bibitem{Qian:2018wid}
X.~Qian and J.~C.~Peng,
``Physics with Reactor Neutrinos,''
Rept. Prog. Phys. \textbf{82}, no.3, 036201 (2019)
[arXiv:1801.05386 [hep-ex]].

\bibitem{DeSalas:2018rby}
P.~F.~De Salas, S.~Gariazzo, O.~Mena, C.~A.~Ternes and M.~T\'ortola,
``Neutrino Mass Ordering from Oscillations and Beyond: 2018 Status and Future Prospects,''
Front. Astron. Space Sci. \textbf{5}, 36 (2018)
[arXiv:1806.11051 [hep-ph]].


\bibitem{Antonelli:2020uui}
V.~Antonelli, L.~Miramonti and G.~Ranucci,
``Present and Future Contributions of Reactor Experiments to Mass Ordering and Neutrino Oscillation Studies,''
Universe \textbf{6}, no.4, 52 (2020)



\bibitem{CHANDLER:2022gvg}
O.~A.~Akindele \textit{et al.} [CHANDLER, CONNIE, CONUS, Daya Bay, JUNO, MTAS, NEOS, NuLat, PROSPECT, RENO, Ricochet, ROADSTR Near-Field Working Group, SoLid, Stereo, Valencia-Nantes TAGS, vIOLETA and WATCHMAN],
``Particle physics using reactor antineutrinos,''
J. Phys. G \textbf{51}, no.8, 080501 (2024)
[arXiv:2203.07214 [hep-ex]].

\bibitem{Esteban:2016qun}
I.~Esteban, M.~C.~Gonzalez-Garcia, M.~Maltoni, I.~Martinez-Soler and T.~Schwetz,
``Updated fit to three neutrino mixing: exploring the accelerator-reactor complementarity,''
JHEP \textbf{01}, 087 (2017)
[arXiv:1611.01514 [hep-ph]].

\bibitem{Forero:2021lax}
D.~V.~Forero, S.~J.~Parke, C.~A.~Ternes and R.~Z.~Funchal,
``JUNO\textquoteright{}s prospects for determining the neutrino mass ordering,''
Phys. Rev. D \textbf{104}, no.11, 113004 (2021)
[arXiv:2107.12410 [hep-ph]].

\bibitem{Parke:2024xre}
S.~J.~Parke and R.~Zukanovich-Funchal,
``A Mass Ordering Sum Rule for the Neutrino Disappearance Channels in T2K, NOvA and JUNO,''
[arXiv:2404.08733 [hep-ph]].


\bibitem{Capozzi:2018dat}
F.~Capozzi, S.~W.~Li, G.~Zhu and J.~F.~Beacom,
``DUNE as the Next-Generation Solar Neutrino Experiment,''
Phys. Rev. Lett. \textbf{123}, no.13, 131803 (2019)
[arXiv:1808.08232 [hep-ph]].



\bibitem{Denton:2023zwa}
P.~B.~Denton and J.~Gehrlein,
``Solar parameters in long-baseline accelerator neutrino oscillations,''
JHEP \textbf{06}, 090 (2023)
[arXiv:2302.08513 [hep-ph]].

\bibitem{Fogli:2004as}
G.~L.~Fogli, E.~Lisi, A.~Marrone, A.~Melchiorri, A.~Palazzo, P.~Serra and J.~Silk,
``Observables sensitive to absolute neutrino masses: Constraints and correlations from world neutrino data,''
Phys. Rev. D \textbf{70}, 113003 (2004)
[arXiv:hep-ph/0408045 [hep-ph]].


\bibitem{Katrin:2024tvg}
M.~Aker \textit{et al.} [Katrin],
``Direct neutrino-mass measurement based on 259 days of KATRIN data,''
[arXiv:2406.13516 [nucl-ex]].

\bibitem{PrivateKATRIN}
A.~Lokhov, private communication.

\bibitem{Gomez-Cadenas:2023vca}
J.~J.~G\'omez-Cadenas, J.~Mart\'\i{}n-Albo, J.~Men\'endez, M.~Mezzetto, F.~Monrabal and M.~Sorel,
``The search for neutrinoless double-beta decay,''
Riv. Nuovo Cim. \textbf{46}, no.10, 619-692 (2023).

\bibitem{Agostini:2022zub}
M.~Agostini, G.~Benato, J.~A.~Detwiler, J.~Men\'endez and F.~Vissani,
``Toward the discovery of matter creation with neutrinoless \ensuremath{\beta}\ensuremath{\beta} decay,''
Rev. Mod. Phys. \textbf{95}, no.2, 025002 (2023)
[arXiv:2202.01787 [hep-ex]].




\bibitem{Agostini:2020xta}
M.~Agostini \textit{et al.} [GERDA],
``Final Results of GERDA on the Search for Neutrinoless Double-$\beta$ Decay,''
Phys. Rev. Lett. \textbf{125}, no.25, 252502 (2020)
[arXiv:2009.06079 [nucl-ex]].


\bibitem{Majorana:2022udl}
I.~J.~Arnquist \textit{et al.} [Majorana],
``Final Result of the Majorana Demonstrator\textquoteright{}s Search for Neutrinoless Double-\ensuremath{\beta} Decay in Ge76,''
Phys. Rev. Lett. \textbf{130}, no.6, 062501 (2023)
[arXiv:2207.07638 [nucl-ex]].

\bibitem{CUORE:2024ikf}
D.~Q.~Adams \textit{et al.} [CUORE],
``With or without $\nu$? Hunting for the seed of the matter-antimatter asymmetry,''
[arXiv:2404.04453 [nucl-ex]].

\bibitem{KamLAND-Zen:2024eml}
S.~Abe \textit{et al.} [KamLAND-Zen],
``Search for Majorana Neutrinos with the Complete KamLAND-Zen Dataset,''
[arXiv:2406.11438 [hep-ex]].

\bibitem{Anton:2019wmi}
G.~Anton \textit{et al.} [EXO-200],
``Search for Neutrinoless Double-$\beta$ Decay with the Complete EXO-200 Dataset,''
Phys. Rev. Lett. \textbf{123}, no.16, 161802 (2019)
[arXiv:1906.02723 [hep-ex]].

\bibitem{Lisi:2022nka}
E.~Lisi and A.~Marrone,
``Majorana neutrino mass constraints in the landscape of nuclear matrix elements,''
Phys. Rev. D \textbf{106}, no.1, 013009 (2022)
[arXiv:2204.09569 [hep-ph]].

\bibitem{Faessler:2008xj}
A.~Faessler, G.~L.~Fogli, E.~Lisi, V.~Rodin, A.~M.~Rotunno and F.~Simkovic,
``QRPA uncertainties and their correlations in the analysis of $0\nu\beta\beta$ decay,''
Phys. Rev. D \textbf{79}, 053001 (2009)
[arXiv:0810.5733 [hep-ph]].

\bibitem{Capozzi:2020}
F.~Capozzi, E.~Di Valentino, E.~Lisi, A.~Marrone, A.~Melchiorri and A.~Palazzo,
``Addendum to: Global constraints on absolute neutrino masses and their ordering,''
Phys. Rev. D \textbf{101}, no.11, 116013 (2020)
[arXiv:2003.08511 [hep-ph]].

\bibitem{Capozzi:2017ipn}
F.~Capozzi, E.~Di Valentino, E.~Lisi, A.~Marrone, A.~Melchiorri and A.~Palazzo,
``Global constraints on absolute neutrino masses and their ordering,''
Phys. Rev. D \textbf{95}, no.9, 096014 (2017)
[arXiv:2003.08511 [hep-ph]].


%
%



%
%








\bibitem{Planck:2018vyg}
N.~Aghanim \textit{et al.} [Planck],
``Planck 2018 results. VI. Cosmological parameters,''
Astron. Astrophys. \textbf{641}, A6 (2020)
[erratum: Astron. Astrophys. \textbf{652}, C4 (2021)]
[arXiv:1807.06209 [astro-ph.CO]].

\bibitem{Palanque-Delabrouille:2019iyz}
N.~Palanque-Delabrouille, C.~Y\`eche, N.~Sch\"oneberg, J.~Lesgourgues, M.~Walther, S.~Chabanier and E.~Armengaud,
``Hints, neutrino bounds and WDM constraints from SDSS DR14 Lyman-$\alpha$ and Planck full-survey data,''
JCAP \textbf{04}, 038 (2020)
[arXiv:1911.09073 [astro-ph.CO]].

\bibitem{DiValentino:2021hoh}
E.~Di Valentino, S.~Gariazzo and O.~Mena,
``Most constraining cosmological neutrino mass bounds,''
Phys. Rev. D \textbf{104}, no.8, 083504 (2021)
[arXiv:2106.15267 [astro-ph.CO]].

\bibitem{Brieden:2022lsd}
S.~Brieden, H.~Gil-Mar\'\i{}n and L.~Verde,
``Model-agnostic interpretation of 10 billion years of cosmic evolution traced by BOSS and eBOSS data,''
JCAP \textbf{08}, no.08, 024 (2022)
[arXiv:2204.11868 [astro-ph.CO]].


\bibitem{DESI:2024mwx}
A.~G.~Adame \textit{et al.} [DESI],
``DESI 2024 VI: Cosmological Constraints from the Measurements of Baryon Acoustic Oscillations,''
[arXiv:2404.03002 [astro-ph.CO]].


\bibitem{DESI:2024uvr}
A.~G.~Adame \textit{et al.} [DESI],
``DESI 2024 III: Baryon Acoustic Oscillations from Galaxies and Quasars,''
[arXiv:2404.03000 [astro-ph.CO]].

\bibitem{DESI:2024kob}
K.~Lodha \textit{et al.} [DESI],
``DESI 2024: Constraints on physics-focused aspects of dark energy using DESI DR1 BAO data,''
Phys. Rev. D \textbf{111}, no.2, 023532 (2025)
[arXiv:2405.13588 [astro-ph.CO]].

\bibitem{DESI:2024lzq}
A.~G.~Adame \textit{et al.} [DESI],
``DESI 2024 IV: Baryon Acoustic Oscillations from the Lyman alpha forest,''
JCAP \textbf{01}, 124 (2025)
[arXiv:2404.03001 [astro-ph.CO]].

\bibitem{DESI:2024aqx}
R.~Calderon \textit{et al.} [DESI],
``DESI 2024: reconstructing dark energy using crossing statistics with DESI DR1 BAO data,''
JCAP \textbf{10}, 048 (2024)
[arXiv:2405.04216 [astro-ph.CO]].

\bibitem{ACT:2020gnv}
S.~Aiola \textit{et al.} [ACT],
``The Atacama Cosmology Telescope: DR4 Maps and Cosmological Parameters,''
JCAP \textbf{12}, 047 (2020)
[arXiv:2007.07288 [astro-ph.CO]].

\bibitem{ACT:2023kun}
M.~S.~Madhavacheril \textit{et al.} [ACT],
``The Atacama Cosmology Telescope: DR6 Gravitational Lensing Map and Cosmological Parameters,''
Astrophys. J. \textbf{962}, no.2, 113 (2024)
[arXiv:2304.05203 [astro-ph.CO]].

\bibitem{ACT:2023dou}
F.~J.~Qu \textit{et al.} [ACT],
``The Atacama Cosmology Telescope: A Measurement of the DR6 CMB Lensing Power Spectrum and Its Implications for Structure Growth,''
Astrophys. J. \textbf{962}, no.2, 112 (2024)
[arXiv:2304.05202 [astro-ph.CO]].

\bibitem{Elbers:2024sha}
W.~Elbers, C.~S.~Frenk, A.~Jenkins, B.~Li and S.~Pascoli,
``Negative neutrino masses as a mirage of dark energy,''
[arXiv:2407.10965 [astro-ph.CO]].

\bibitem{Naredo-Tuero:2024sgf}
D.~Naredo-Tuero, M.~Escudero, E.~Fern\'andez-Mart\'\i{}nez, X.~Marcano and V.~Poulin,
``Critical look at the cosmological neutrino mass bound,''
Phys. Rev. D \textbf{110}, no.12, 123537 (2024)
[arXiv:2407.13831 [astro-ph.CO]].

\bibitem{Craig:2024tky}
N.~Craig, D.~Green, J.~Meyers and S.~Rajendran,
``No \ensuremath{\nu}s is Good News,''
JHEP \textbf{09}, 097 (2024)
[arXiv:2405.00836 [astro-ph.CO]].

\bibitem{Green:2024xbb}
D.~Green and J.~Meyers,
``The Cosmological Preference for Negative Neutrino Mass,''
[arXiv:2407.07878 [astro-ph.CO]].









\bibitem{Herold:2024nvk}
L.~Herold and M.~Kamionkowski,
``Revisiting the impact of neutrino mass hierarchies on neutrino mass constraints in light of recent DESI data,''
[arXiv:2412.03546 [astro-ph.CO]].

\bibitem{Escudero:2024uea}
H.~G.~Escudero and K.~N.~Abazajian,
``The Status of Neutrino Cosmology: Standard $\Lambda$CDM, Extensions, and Tensions,''
[arXiv:2412.05451 [astro-ph.CO]].

\bibitem{RoyChoudhury:2024wri}
S.~Roy Choudhury and T.~Okumura,
``Updated Cosmological Constraints in Extended Parameter Space with Planck PR4, DESI Baryon Acoustic Oscillations, and Supernovae: Dynamical Dark Energy, Neutrino Masses, Lensing Anomaly, and the Hubble Tension,''
Astrophys. J. Lett. \textbf{976}, no.1, L11 (2024)
[arXiv:2409.13022 [astro-ph.CO]].


\bibitem{DESI:2024hhd}
A.~G.~Adame \textit{et al.} [DESI],
``DESI 2024 VII: Cosmological Constraints from the Full-Shape Modeling of Clustering Measurements,''
[arXiv:2411.12022 [astro-ph.CO]].

\bibitem{Reboucas:2024smm}
J.~Rebou\c{c}as, D.~H.~F.~de Souza, K.~Zhong, V.~Miranda and R.~Rosenfeld,
``Investigating Late-Time Dark Energy and Massive Neutrinos in Light of DESI Y1 BAO,''
[arXiv:2408.14628 [astro-ph.CO]].

\bibitem{Upadhye:2024ypg}
A.~Upadhye, M.~R.~Mosbech, G.~Pierobon and Y.~Y.~Y.~Wong,
``Everything hot everywhere all at once: Neutrinos and hot dark matter as a single effective species,''
[arXiv:2410.05815 [astro-ph.CO]].

\bibitem{Shao:2024mag}
H.~Shao, J.~J.~Givans, J.~Dunkley, M.~Madhavacheril, F.~Qu, G.~Farren and B.~Sherwin,
``Cosmological limits on the neutrino mass sum for beyond-$\Lambda$CDM models,''
[arXiv:2409.02295 [astro-ph.CO]].


\bibitem{Allali:2024aiv}
I.~J.~Allali and A.~Notari,
``Neutrino mass bounds from DESI 2024 are relaxed by Planck PR4 and cosmological supernovae,''
JCAP \textbf{12}, 020 (2024)
[arXiv:2406.14554 [astro-ph.CO]].





\bibitem{Wang:2024pui}
Z.~Wang, S.~Lin, Z.~Ding and B.~Hu,
``The role of LRG1 and LRG2\textquoteright{}s monopole in inferring the DESI 2024 BAO cosmology,''
Mon. Not. Roy. Astron. Soc. \textbf{534}, no.4, 3869-3875 (2024)
[arXiv:2405.02168 [astro-ph.CO]].

\bibitem{Colgain:2024xqj}
E.~\'O.~Colg\'ain, M.~G.~Dainotti, S.~Capozziello, S.~Pourojaghi, M.~M.~Sheikh-Jabbari and D.~Stojkovic,
``Does DESI 2024 Confirm $\Lambda$CDM?,''
[arXiv:2404.08633 [astro-ph.CO]].


\bibitem{Sapone:2024ltl}
D.~Sapone and S.~Nesseris,
``Outliers in DESI BAO: robustness and cosmological implications,''
[arXiv:2412.01740 [astro-ph.CO]].


\bibitem{Calabrese:2008rt}
E.~Calabrese, A.~Slosar, A.~Melchiorri, G.~F.~Smoot and O.~Zahn,
``Cosmic Microwave Weak lensing data as a test for the dark universe,''
Phys. Rev. D \textbf{77}, 123531 (2008)
[arXiv:0803.2309 [astro-ph]].


\bibitem{DiValentino:2015bja}
E.~Di Valentino, A.~Melchiorri and J.~Silk,
``Cosmological hints of modified gravity?,''
Phys. Rev. D \textbf{93}, no.2, 023513 (2016)
[arXiv:1509.07501 [astro-ph.CO]].


\bibitem{Renzi:2017cbg}
F.~Renzi, E.~Di Valentino and A.~Melchiorri,
``Cornering the Planck $A_{lens}$ anomaly with future CMB data,''
Phys. Rev. D \textbf{97}, no.12, 123534 (2018)
[arXiv:1712.08758 [astro-ph.CO]].


\bibitem{Domenech:2020qay}
G.~Dom\`enech, X.~Chen, M.~Kamionkowski and A.~Loeb,
``Planck residuals anomaly as a fingerprint of alternative scenarios to inflation,''
JCAP \textbf{10}, 005 (2020)
[arXiv:2005.08998 [astro-ph.CO]].


\bibitem{Rosenberg:2022sdy}
E.~Rosenberg, S.~Gratton and G.~Efstathiou,
``CMB power spectra and cosmological parameters from Planck PR4 with CamSpec,''
Mon. Not. Roy. Astron. Soc. \textbf{517}, no.3, 4620-4636 (2022)
[arXiv:2205.10869 [astro-ph.CO]].


\bibitem{Tristram:2023haj}
M.~Tristram, A.~J.~Banday, M.~Douspis, X.~Garrido, K.~M.~G\'orski, S.~Henrot-Versill\'e, L.~T.~Hergt, S.~Ili\'c, R.~Keskitalo and G.~Lagache, \textit{et al.}
``Cosmological parameters derived from the final Planck data release (PR4),''
Astron. Astrophys. \textbf{682}, A37 (2024)
[arXiv:2309.10034 [astro-ph.CO]].


\bibitem{Scolnic:2021amr}
D.~Scolnic, D.~Brout, A.~Carr, A.~G.~Riess, T.~M.~Davis, A.~Dwomoh, D.~O.~Jones, N.~Ali, P.~Charvu and R.~Chen, \textit{et al.}
``The Pantheon+ Analysis: The Full Data Set and Light-curve Release,''
Astrophys. J. \textbf{938}, no.2, 113 (2022)
[arXiv:2112.03863 [astro-ph.CO]].


\bibitem{Brout:2022vxf}
D.~Brout, D.~Scolnic, B.~Popovic, A.~G.~Riess, J.~Zuntz, R.~Kessler, A.~Carr, T.~M.~Davis, S.~Hinton and D.~Jones, \textit{et al.}
``The Pantheon+ Analysis: Cosmological Constraints,''
Astrophys. J. \textbf{938}, no.2, 110 (2022)
[arXiv:2202.04077 [astro-ph.CO]].


\bibitem{Rubin:2023ovl}
D.~Rubin, G.~Aldering, M.~Betoule, A.~Fruchter, X.~Huang, A.~G.~Kim, C.~Lidman, E.~Linder, S.~Perlmutter and P.~Ruiz-Lapuente, \textit{et al.}
``Union Through UNITY: Cosmology with 2,000 SNe Using a Unified Bayesian Framework,''
[arXiv:2311.12098 [astro-ph.CO]].


\bibitem{DES:2024tys}
T.~M.~C.~Abbott \textit{et al.} [DES],
``The Dark Energy Survey: Cosmology Results with \ensuremath{\sim}1500 New High-redshift Type Ia Supernovae Using the Full 5 yr Data Set,''
Astrophys. J. Lett. \textbf{973}, no.1, L14 (2024)
[arXiv:2401.02929 [astro-ph.CO]].


\bibitem{DES:2024upw}
B.~O.~S\'anchez \textit{et al.} [DES],
``The Dark Energy Survey Supernova Program: Light Curves and 5 Yr Data Release,''
Astrophys. J. \textbf{975}, no.1, 5 (2024)
[arXiv:2406.05046 [astro-ph.CO]].


\bibitem{DES:2024hip}
M.~Vincenzi \textit{et al.} [DES],
``The Dark Energy Survey Supernova Program: Cosmological Analysis and Systematic Uncertainties,''
Astrophys. J. \textbf{975}, no.1, 86 (2024)
[arXiv:2401.02945 [astro-ph.CO]].


\bibitem{Carron:2022eyg}
J.~Carron, M.~Mirmelstein and A.~Lewis,
``CMB lensing from Planck PR4~maps,''
JCAP \textbf{09}, 039 (2022)
[arXiv:2206.07773 [astro-ph.CO]].


\bibitem{Planck:2020olo}
Y.~Akrami \textit{et al.} [Planck],
``$Planck$ intermediate results. LVII. Joint Planck LFI and HFI data processing,''
Astron. Astrophys. \textbf{643}, A42 (2020)
[arXiv:2007.04997 [astro-ph.CO]].


\bibitem{Notari:2024zmi}
A.~Notari, M.~Redi and A.~Tesi,
``BAO vs. SN evidence for evolving dark energy,''
[arXiv:2411.11685 [astro-ph.CO]].


\bibitem{Giare:2025pzu}
W.~Giar\`e, T.~Mahassen, E.~Di Valentino and S.~Pan,
``An overview of what current data can (and cannot yet) say about evolving dark energy,''
[arXiv:2502.10264 [astro-ph.CO]].


\bibitem{Efstathiou:2024xcq}
G.~Efstathiou,
``Evolving Dark Energy or Supernovae Systematics?,''
[arXiv:2408.07175 [astro-ph.CO]].


\bibitem{Lewis:1999bs}
A.~Lewis, A.~Challinor and A.~Lasenby,
``Efficient computation of CMB anisotropies in closed FRW models,''
Astrophys. J. \textbf{538}, 473-476 (2000)
[arXiv:astro-ph/9911177 [astro-ph]].


\bibitem{Howlett:2012mh}
C.~Howlett, A.~Lewis, A.~Hall and A.~Challinor,
``CMB power spectrum parameter degeneracies in the era of precision cosmology,''
JCAP \textbf{04}, 027 (2012)
[arXiv:1201.3654 [astro-ph.CO]].


\bibitem{Lewis:2002ah}
A.~Lewis and S.~Bridle,
``Cosmological parameters from CMB and other data: A Monte Carlo approach,''
Phys. Rev. D \textbf{66}, 103511 (2002)
[arXiv:astro-ph/0205436 [astro-ph]].


\bibitem{Lewis:2013hha}
A.~Lewis,
``Efficient sampling of fast and slow cosmological parameters,''
Phys. Rev. D \textbf{87}, no.10, 103529 (2013)
[arXiv:1304.4473 [astro-ph.CO]].


\bibitem{Colgain:2024mtg}
E.~\'O.~Colg\'ain and M.~M.~Sheikh-Jabbari,
``DESI and SNe: Dynamical Dark Energy, $\Omega_m$ Tension or Systematics?,''
[arXiv:2412.12905 [astro-ph.CO]].



\bibitem{CERNnuWeek}
Contributions to the Workshop ``CERN Neutrino Platform Pheno Week'' (CERN, Geneva, Feb.~2025), website: https://indico.cern.ch/event/1454726/

\bibitem{Neutel2023}
E.\ Lisi, ``Towards a global analysis of absolute neutrino masses,'' in {\em NeuTel 2023\/},  XX International Workshop on Neutrino Telescopes (Venice, Italy, Oct.~2023). Website: https://agenda.infn.it/event/33107/

\end{thebibliography}
\end{document}